\documentclass[fleqn,usenatbib]{mnras}

\usepackage{amsmath}
\usepackage{mathtools}
\usepackage[normalem]{ulem} 
\usepackage{natbib}
\usepackage{comment}
\usepackage{soul} 
\usepackage{xspace}
\usepackage{xifthen}
\usepackage{lineno}

\usepackage[T1]{fontenc}

\usepackage{soul} 
\usepackage{amsmath}
\usepackage{amssymb}
\usepackage{xspace}
\usepackage{xifthen}

\newcommand{\unit}[1]{\ensuremath{\mathrm{\,#1}}\xspace}

\newcommand{\mas}{\unit{mas}}

\newcommand{\km}{\unit{km}}
\newcommand{\kms}{\km \second^{-1}}
\newcommand{\pc}{\unit{pc}}
\newcommand{\kpc}{\unit{kpc}}
\newcommand{\second}{\unit{s}}

\newcommand{\Msun}{\unit{M_\odot}}

\newcommand{\yr}{\unit{yr}}
\newcommand{\masyr}{\unit{\mas \yr^{-1}}}

\newcommand{\COMMENT}[3]{{}}

\newcommand{\NEW}{\textcolor{blue}}

\newcommand{\kgotwo}[0]{Gran~3\xspace}
\newcommand{\kgofour}[0]{Gran~4\xspace}
\newcommand{\kgoseven}[0]{Gaia~9\xspace}
\newcommand{\kgoeight}[0]{LP~866\xspace}
\newcommand{\kgoten}[0]{Gaia~10\xspace}
\newcommand{\kgotwotwo}[0]{Garro~01\xspace}

\usepackage{newtxtext,newtxmath}

\DeclareRobustCommand{\VAN}[3]{#2}
\let\VANthebibliography\thebibliography
\def\thebibliography{\DeclareRobustCommand{\VAN}[3]{##3}\VANthebibliography}

\usepackage{graphicx}	
\usepackage{amsmath}	

\title[Magellan/M2FS Spectroscopy of Six Star Clusters]{ The Kinematics, Metallicities, and Orbits of Six Recently Discovered  Galactic Star Clusters with  Magellan/M2FS Spectroscopy\thanks{This paper presents data gathered with the Magellan Telescopes at Las Campanas Observatory, Chile.}}

\author[Andrew B. Pace et al.]{
Andrew B. Pace,$^{1}$\thanks{E-mail: apace@andrew.cmu.edu}
Sergey E. Koposov,$^{2, 3, 4, 1}$
Matthew G. Walker,$^{1}$
Nelson Caldwell,$^{5}$
Mario Mateo, $^{6}$ 
\newauthor
Edward W. Olszewski, $^{7}$
Ian U. Roederer, $^{6, 8}$
John I. Bailey, III, $^{9}$
Vasily Belokurov,$^{3}$
Kyler Kuehn, $^{10}$
\newauthor
Ting S. Li, $^{11}$
Daniel B. Zucker $^{12,13}$
\\
$^{1}$McWilliams Center for Cosmology, Carnegie Mellon University, 5000 Forbes Ave, Pittsburgh, PA 15213, USA\\
$^{2}$Institute for Astronomy, University of Edinburgh, Royal Observatory, Blackford Hill, Edinburgh EH9 3HJ, UK\\
$^{3}$Institute of Astronomy, University of Cambridge, Madingley Road, Cambridge CB3 0HA, UK\\
$^{4}$Kavli Institute for Cosmology, University of Cambridge, Madingley Road, Cambridge CB3 0HA, UK\\
$^{5}$Harvard-Smithsonian Center for Astrophysics, 60 Garden Street, MS-15, Cambridge, MA 02138, USA\\
$^{6}$Department of Astronomy, University of Michigan, Ann Arbor, MI 48109, USA\\
$^{7}$Steward Observatory, The University of Arizona, 933 N. Cherry Avenue, Tucson, AZ 85721, USA\\
$^{8}$Joint Institute for Nuclear Astrophysics – Center for the Evolution of the Elements (JINA-CEE), USA\\
$^{9}$Department of Physics, UCSB, Santa Barbara, CA 93016, USA\\
$^{10}$ Lowell Observatory, 1400 W Mars Hill Rd, Flagstaff,  AZ 86001, USA\\
$^{11}$Department of Astronomy and Astrophysics, University of Toronto, 50 St. George Street, Toronto ON, M5S 3H4, Canada\\
$^{12}$ School of Mathematical and Physical Sciences, Macquarie University, Sydney, NSW 2109, Australia\\
$^{13}$ Macquarie University Research Centre for Astronomy, Astrophysics \& Astrophotonics, Sydney, NSW 2109, Australia\\
}

\date{Accepted XXX. Received YYY; in original form ZZZ}

\pubyear{2023}

\begin{document}
\label{firstpage}
\pagerange{\pageref{firstpage}--\pageref{lastpage}}
\maketitle

\begin{abstract}
We present Magellan/M2FS  spectroscopy of four recently discovered Milky Way star clusters (\kgotwo/Patchick~125, \kgofour, \kgotwotwo, \kgoeight) and two newly discovered open clusters (\kgoseven, \kgoten) at low Galactic latitudes. 
We measure line-of-sight velocities and  stellar parameters ([Fe/H], $\log{g}$, $T_{\rm eff}$, [Mg/Fe]) from high resolution spectroscopy centered on the Mg triplet and identify 20-80 members per star cluster. 
We determine the kinematics and chemical properties of each cluster and  measure the systemic proper motion and orbital properties by utilizing  {\it Gaia} astrometry. 
We find \kgotwo to  be an old, metal-poor (mean metallicity of ${\rm [Fe/H]}=-1.83$) globular cluster located in the Galactic bulge on  a retrograde orbit.
\kgofour is an old, metal-poor (${\rm [Fe/H]}=-1.84$) globular cluster with a halo-like orbit that happens to be  passing through the Galactic plane. 
The orbital properties of \kgofour are consistent with the proposed  LMS-1/Wukong and/or Helmi streams merger events. 
\kgotwotwo is   metal-rich (${\rm [Fe/H]}=-0.30$) and on a near circular orbit in the outer disk but its classification as an  open  cluster or globular cluster is ambiguous. 
\kgoseven and \kgoten  are among the most distant known open clusters  
 at $R_{GC}\sim18,~21.2~\kpc$ and most metal-poor with [Fe/H] $\sim-0.50,-0.34$ for \kgoseven and \kgoten, respectively. 
\kgoeight is a nearby, metal-rich open cluster ([Fe/H]$=+0.10$).
The  discovery and confirmation of multiple  star clusters in the Galactic plane shows the power of {\it Gaia} astrometry and the star cluster census  remains incomplete.
\end{abstract}




\section{Introduction}

\label{sec:intro}

Star clusters are among the smallest stellar structures in the universe and are a key component of hierarchical structure assembly. 
They are valuable for studying  stellar populations and their evolution at a variety of ages,  metallicities, and environs \citep[e.g.,][]{Krumholz2019ARA&A..57..227K, Adamo2020SSRv..216...69A}. 
Star clusters in the Milky Way (MW) are typically divided into two categories: the older, denser, and more luminous globular clusters \citep{Gratton2019A&ARv..27....8G}, and the younger clusters in the MW disk, referred to as open clusters \citep[e.g.,][]{CantatGaudin2022Univ....8..111C}.

While the census of bright halo clusters is mostly complete \citep{Webb2021MNRAS.502.4547W}, there has been a number of faint star clusters discovered in optical wide-field imaging surveys, pushing the luminosity and surface brightness boundary \citep[e.g,][]{Koposov2007ApJ...669..337K, Belokurov2014MNRAS.441.2124B, Torrealba2019MNRAS.484.2181T, Mau2019ApJ...875..154M, Cerny2023ApJ...953....1C}.
The census of star clusters in the MW mid-plane is  incomplete due to the high extinction and large stellar foreground.
With recent near-infrared surveys such as the VISTA Variables in the Via L\'{a}ctea Survey (VVV)
\citep[e.g.,][]{Minniti2011A&A...527A..81M, Garro2020A&A...642L..19G} and 
astrometric data from the {\it Gaia} mission the number of star cluster candidates has significantly increased  \citep[e.g.,][]{Koposov2017MNRAS.470.2702K, Torrealba2019MNRAS.484.2181T, Garro2020A&A...642L..19G, Gran2022MNRAS.509.4962G}. 
However, a number of the star cluster candidates found pre-{\it Gaia} DR2 have been shown to be false positives once proper motions and kinematics are considered \citep{Gran2019A&A...628A..45G, CantatGaudin2020A&A...633A..99C}.

Ultimately, stellar spectroscopy is required to validate   candidate star clusters   and confirm they are not a mirage of MW stars \citep[e.g.,][]{Gran2022MNRAS.509.4962G}. Furthermore, spectroscopic radial velocities and metallicities will identify star cluster members.  With the systemic  radial velocity, the orbit and origin of a star cluster can be determined \citep[e.g.,][]{Massari2019A&A...630L...4M, Kruijssen2019MNRAS.486.3180K}and the internal dynamics analyzed with large radial velocity samples \citep[e.g.,][]{Baumgardt2018MNRAS.478.1520B, Garro2023A&A...669A.136G}. 
Spectroscopic metallicities can assist in determining the classification and origin of a star cluster \citep[e.g.,][]{Gran2022MNRAS.509.4962G}. 

We present spectroscopic confirmation of three recently discovered globular cluster candidates (\kgotwo, \kgofour, \kgotwotwo) and three newly discovered open clusters (\kgoseven, \kgoten, \kgoeight). 
In Section~\ref{section:discovery}, we discuss our search algorithm and independent discovery of the star cluster candidates with {\it Gaia} DR2. 
In Section~\ref{section:data}, we discuss our spectroscopic observations, velocity and metallicity measurements, and the auxiliary data analyzed.
In Section~\ref{label:results}, we identify members of each star cluster,  measure the general kinematic and metallicity properties, measure the spatial distribution, and determine  the orbital properties. 
In Section~\ref{section:discussion}, we analyze the globular cluster internal kinematics, compare the globular clusters to other  MW globular clusters, discuss the origin and potential association to accretion events, analyze the open clusters in the context of the Galactic metallicity gradient, and compare our results to the literature. We  summarize our conclusions in Section~\ref{section:conclusion}.  

\section{Discovery and Candidate identification}
\label{section:discovery}

The search for the stellar overdensities was carried out in 2018 after the release of the {\it Gaia} DR2 using the satellite detection pipeline broadly based on the methods presented in  \citet{Koposov2008ApJ...686..279K, Koposov2015ApJ...805..130K}, but extended into space of proper motions. 

We describe here briefly the basics behind the detection algorithm, leaving a more detailed description to a separate contribution (Koposov et al. in prep.). 

The algorithm consists of several steps.
\begin{itemize}
    \item Looping over all proper motions.  In  the search used here we ran overdensity search for subsets of stars with proper motions $|\mu_\alpha-X_i|<1$, $|\mu_\delta-X_j|<1$ where $X_i$,$X_j$ span the range of proper motions from -15 to 15 mas/year with 1 mas/year steps.
    \item Looping over all possible distance moduli to overdensities. We perform the search for stars selected based on an isochrone filter placed at distances from $\sim$ 6\,kpc to 160\,kpc. We used the extinction corrected BP, RP and G magnitudes and an old metal-poor PARSEC isochrone \citep{Bressan2012MNRAS.427..127B} with an age of 12 Gyr  and  $[{\rm Fe/H}]=-2$. We also run a single search without any isochrone colour magnitude selection.
    \item Looping over overdensity sizes from 3 arcmin to 48 arcmin.
    \item Segmentation of the sky into HEALPIX \citep{Gorski2011} tiles. To avoid having to work with the dataset for the entire sky the overdensity search algorithm works with approximately rectangular-shaped HEALPIX tiles, that we also increase in size by 20\% with respect to the standard HEALPIX scheme to ensure overlap between tiles and avoid dealing with edge effects. The exact Nside resolution parameter and pixel scale of tiling were different for different runs depending on  memory limitations and size of the overdensity being searched for.
    \item Creation of a pixelated stellar density map inside each tile. We use tangential projection to map the stars into a rectangular x,y pixel grid, and then make a 2-D histogram of stellar counts of stars selected by proper motions, colours and magnitudes. 
    \item Creation of a stellar overdensity significance map based on a stellar count map.  This step is described below in more detail.
    \item Identification of overdensities based on the significance map and merging of candidate lists from various search configurations.
    \item Cross-matching the candidate lists with external catalogues and construction of validation plots.
\end{itemize}

Below we provide a brief description of the algorithm that provides an overdensity significance map given the binned 2D stellar density map. The algorithm requires three main parameters -- the kernel size, which corresponds to the size of the overdensity $k$ we are looking for, and two background apertures, $b_1$ and $b_2$. 
Here we assume that we have a rectangular grid of stellar number counts $H(x,y)$ and we estimate the significance of the overdensity at pixel x=0, y=0. The key difference of our approach compared to previous approaches  \citep[i.e.,][]{Koposov2008ApJ...686..279K} is that we do not rely on the assumption of Gaussianity or even a Poisson distribution of number counts in the map.

We first compute the number of stars $N$ in a circular aperture with radius k around the pixel 0,0. We then need to characterize what is the probability distribution of $P_{null}(N)$ under a null hypothesis of no overdensity to compute the tail probability/significance. 

Our model for $P_{null}(N)$ is the negative-binomial distribution, which is a discrete Poisson-like distribution (that is, an infinite mixture of Poisson distributions with means having a Gamma distribution). The negative binomial distribution can be parameterized with mean $\mu$ and $\sigma^2$, where $\sigma^2\ge \mu$ (note that the variance of the negative-binomial distribution is a parameter, as opposed to a Poisson distribution, where it is equal to the mean). The $\mu$ and $\sigma$ are estimated based on the number count distribution between two background apertures $b_1$ and $b_2$.

The significance (or the Z-score) in each pixel is then assigned as $Z=F^{-1}(P_{null}(\geq N))$, where $F^{-1}()$ is the inverse of the CDF of a normal distribution. The significant overdensities are selected as those where Z is larger than a certain threshold. For this work we used $Z>6$ selected candidates.

The application of the algorithm summarized above to {\it Gaia} DR2 in 2018 yielded a few hundred significant distinct overdensities. The absolute majority of them were known, but around 30 objects were deemed to be likely real dwarf galaxies or globular clusters and were selected for further inspection. The spectroscopic follow-up of six of these objects is the subject of this paper. Some objects from the list, such as the Eridanus IV object \citep{Cerny2021ApJ...920L..44C} have been discovered and independently followed up since.

We began our spectroscopic follow-up in 2018 and note that 4 star clusters in our sample  have since been independently discovered.  We refer to these clusters by their name in the first discovery analysis. \kgotwotwo was independently discovered by  \citet{Garro2020A&A...642L..19G} in the near-IR VISTA Variables in the Via L\'{a}ctea Extended Survey (VVVX).
\citet{Gran2022MNRAS.509.4962G} independently discovered \kgotwo (also known as Patchick~125) and \kgofour with {\it Gaia} DR2 astrometry, and they were confirmed with the VVV survey.  Gran~3 was independently discovered by the amateur astronomer Dana Patchick and named Patchick~125. 
We used the literature open cluster compilation from \citet{Hunt2023arXiv230313424H} to search for literature cross matches for our open clusters which includes most post-{\it Gaia} open cluster discoveries \citep[e.g.,][]{Bica2019AJ....157...12B, Liu2019ApJS..245...32L, CantatGaudin2020A&A...633A..99C, Kounkel2020AJ....160..279K, CastroGinard2022A&A...661A.118C}.
One cluster, internally KGO~8, was independently discovered by \citet{Liu2019ApJS..245...32L} and referred to as LP~866 (although in some catalogs it is referred to as FoF~866) and \citet{Kounkel2020AJ....160..279K} and referred to as Theia~4124. We refer to this open cluster as \kgoeight here.
The two remaining open clusters in our sample are new discoveries, and we name them \kgoseven and \kgoten.

\section{Spectroscopic Follow-Up}
\label{section:data}

\begin{table*}
	\centering
	\caption{Spectroscopic Observations of  Star Clusters.}
	\label{tab:spectroscopic_observations}
	\begin{tabular}{lll llll l } 
		\hline
 object & R.A. (deg) & Dec. (deg) & Telescope/Instrument & UT Date & Exp. Time  & $N_\mathrm{obs}$ & $N_\mathrm{good}$\\
 \hline
\kgotwo & $256.135833$ & $-35.471278$ & Magellan/M2FS & 2018-08-11 & 6900 & 54 & 41\\
\kgofour & $278.092083$ & $-23.211194$ & Magellan/M2FS & 2018-08-13 & 6100 & 118 & 115\\
\kgofour & $278.112121$ & $-23.103756$ & AAT/AAOmega & 2018-06-24 & 1800 & 67 & 67 \\
\kgotwotwo & $212.246250$& $-65.738333$ & Magellan/M2FS & 2018-08-11 & 5400 & 205 & 193\\
\kgoseven & $119.607083$ & $-38.984639$& Magellan/M2FS & 2018-12-06 & 5400 & 96 & 50 \\
\kgoten & $121.172500$& $-38.984444$& Magellan/M2FS & 2018-08-15 & 5800 & 56 & 43\\
\kgoeight & $261.651250$& $-39.280889$& Magellan/M2FS & 2018-12-05 & 7200 & 164 & 160\\
		\hline
	\end{tabular}
\end{table*}

\subsection{Spectroscopic targeting}

The possible member stars from candidate stellar overdensities discovered 
were selected for spectroscopic observations by  using the information about the objects that was available from their detection, such as approximate object angular size and proper motion. We did not have a uniform target selection strategy from object to object, so we provide a broad overview of the selection. We typically targeted stars using {\it Gaia} DR2 astrometry and photometry, selecting  stars with proper motions within 1-3 mas/yr of the center of the detection. We applied the {\tt astrometric\_excess\_noise} <1 cut and selected stars with small parallaxes $\varpi< {\rm Max}(0.1 , 3\,\sigma_\varpi)$. 
Since the majority of followed-up overdensities had small angular sizes we tried to maximise the number of fibers on each object by assigning higher priority to central targets. We also did not apply any colour-magnitude or isochrone selection masks to the targets, other than a magnitude limit to ensure sufficient signal to noise, and prioritising brighter stars, such as $G<16-18$. 

\subsection{M2FS Spectroscopy}

We present spectroscopic observations of six star cluster candidates that we obtained using the Michigan/Magellan Fiber System \citep[M2FS;][]{Mateo2012SPIE.8446E..4YM} at the 6.5-m Magellan/Clay Telescope at Las Campanas Observatory, Chile.  
M2FS deploys 256 fibers over a field of diameter  $0.5^{\circ}$, feeding two independent spectrographs that offer various modes of configuration.  We used both spectrographs in identical configurations that provide resolving power $\mathcal{R}\sim 24,000$ over the spectral range $5130 - 5190$ \AA.  For all six clusters, Table \ref{tab:spectroscopic_observations} lists coordinates of the M2FS field center, UT date and exposure time of the observation, the number of science targets and the number that yielded `good' observations that pass our quality-control criteria.  

We process and model all M2FS spectra using the procedures described in detail by \citet{Walker2023ApJS..268...19W}.  Briefly, we use custom Python-based software to execute standard processing steps (e.g., overscan, bias and dark corrections), to identify and trace spectral apertures, to extract 1D spectra, to calibrate wavelengths, to correct for variations in pixel sensitivity and fiber throughput, and finally to subtract the mean sky level measured from $\sim 20$ fibers per field that are pointed toward regions of blank sky.  To each individually-processed spectrum, we fit a model based on a library of synthetic template spectra computed on a regular grid of stellar-atmospheric parameters: effective temperature ($T_{\rm eff}$), surface gravity ($\log g$), metallicity ([Fe/H]) and magnesium abundance ([Mg/Fe]).  Including parameters that adjust the resolution and continuum level of the template spectra, our spectral model has 16 free parameters.  We use the software package MultiNest to draw random samples from the 16-dimensional posterior probability distribution function (PDF) \citep{Feroz2008MNRAS.384..449F, Feroz2009MNRAS.398.1601F}.  We summarize 1D posterior PDFs for each of the physical parameters according to the mean and standard deviation of the sample returned by MultiNest.  

We consider stars with ${\rm S/N}>0$ and $\sigma_{v_{\rm los}}>5\kms$ as good quality measurements \citep{Walker2023ApJS..268...19W}. 
We consider stars  with ${\rm S/N}>2$ to be good quality [Fe/H] measurements. 
Our selection of cuts for good quality [Fe/H] is based on repeat measurements and dwarf galaxy data in the M2FS catalog \citep{Walker2023ApJS..268...19W}.

\subsection{AAT Observations}

One of the objects detected in the {\it Gaia} search was submitted for follow-up observations by the Two-degree Field (2dF) spectrograph \citep{Lewis2002MNRAS.333..279L} at the Anglo-Australian Telescope (AAT). The observations were conducted during the observing run of the Southern  Stellar Stream  Spectroscopic Survey ($S^5$) \citep{Li2019MNRAS.490.3508L}. In particular, an internal data release (iDR3.1) was used for this work. We refer to \citet{Li2022ApJ...928...30L} for a detailed description of the data and processing and provide a brief summary here. The stars were observed with two arms of the spectrograph: the red arm  with the 1700D grating that covers a wavelength range from 8400 to 8800 \AA\ (including Ca~\textsc{ii} near-infrared triplet with $\lambda\lambda$8498, 8542, and 8662) with a spectral resolution of $R\sim 10000$, and the blue arm with the 580V grating that provides low resolution $R\sim 1300$ spectra covering a broad wavelength range from 3800 to 5800 \AA. The blue and red spectra for each star were then forward modeled by the {\tt rvspecfit} code \citep{rvspecfit2019} to provide estimates of  stellar parameters, radial velocities and their uncertainties. In addition to the velocities, we also acquired the calcium triplet (CaT) metallicities from equivalent widths and the \citet{Carrera2013MNRAS.434.1681C} calibration for all the member stars as detailed in \citet{Li2022ApJ...928...30L}.

\subsection{Additional Data}
\label{section:gaia}

We use photometric and astrometric data from the {\it Gaia} EDR3 catalog \citep{Gaia_Brown_2021A&A...649A...1G}.
We only utilize astrometric data that passed the following cuts: {\tt ruwe} $<1.5$ \citep{Gaia_Lindegren_2021A&A...649A...2L},   and \texttt{astrometric\_excess\_noise\_sig}$<3$.
We note that some  stars that were targeted spectroscopically do not pass these quality cuts (partly due to the {\it Gaia} DR2 target selection) and we exclude those stars from any   astrometry based analysis. 
For  parallax measurements, we apply the parallax offset from \citet{Gaia_Lindegren_2021A&A...649A...2L} and include an additional offset of $\Delta\varpi=0.007~{\rm mas}$ based on the globular cluster analysis of \citet{Vasiliev2021MNRAS.505.5978V}.

We use  {\it Gaia} DR3  RR Lyrae (RRL) catalog to search for candidate RRL star cluster members \citep{Clementini2022arXiv220606278C}.
We use the  DECam Plane Survey (DECaPS) DR1  griz photometric data for \kgotwo, \kgotwotwo and \kgoeight \citep{Schlafly2018ApJS..234...39S}.
We search for additional spectroscopic members of our star cluster sample in large spectroscopic surveys including SDSS  APOGEE DR17 \citep{SDSS_DR17_2022ApJS..259...35A}, GALAH \citep{Buder2021MNRAS.506..150B}, and  {\it Gaia} RVS DR3 \citep{Katz2022arXiv220605902K}.

\section{Results}
\label{label:results}

\begin{table*}
	\centering
	\caption{Properties of the Star Clusters. Literature $M_V$ measurements of the \kgotwo, \kgofour, and \kgotwotwo are from \citet{Garro2022A&A...659A.155G, Gran2022MNRAS.509.4962G, Garro2020A&A...642L..19G}. }
	\label{tab:kgo_properties}
	\begin{tabular}{l c c c c c c } 
		\hline
 & Gran~3/Patchick~125 & Gran~4 & Garro~01 & Gaia~9 & Gaia~10 & LP~866 \\
 \hline
 \\
R.A. (J2000, deg) & 256.24 & 278.113 & 212.25 & 119.707 & 121.168 & 261.766 \\
Dec (J2000, deg) & -35.49 & -23.105 & -65.62 & -39.011 & -38.928 & -39.439 \\
l (deg) & 349.75 & 10.20 & 310.83 & 254.65 & 255.17 & 349.09 \\
b (deg) & 3.44 & -6.38 & -3.94 & -4.97 & -3.96 & -2.42 \\
$r_h$ (arcmin) & $1.7\pm0.2$ & $2.2_{-0.4}^{+0.5}$ & $2.4_{-0.4}^{+0.6}$ & $1.4\pm0.2$ & $1.6_{-0.2}^{+0.3}$ & $4.6_{-0.6}^{+0.7}$ \\
$r_h$ (parsec) & $5.3_{-0.6}^{+0.7}$ & $14.2_{-2.5}^{+3.3}$ & $10.9_{-2.0}^{+2.6}$ & $5.5_{-0.7}^{+0.9}$ & $7.8_{-1.2}^{+1.6}$ & $3.1_{-0.4}^{+0.5}$ \\
$r_c$ (arcmin) & $1.1_{-0.2}^{+0.3}$ & $1.4_{-0.4}^{+0.5}$ & $1.8_{-0.5}^{+0.7}$ & $0.8\pm0.2$ & $1.0_{-0.2}^{+0.3}$ & $3.3_{-0.7}^{+0.8}$ \\
$r_t$ (arcmin) & $>5.3$ & $>5.8$ & $>5.3$ & $>5.4$ & $>5.0$ & $>11.4$ \\
$D$ (kpc) & $10.5$ & $21.9$ & $15.5$ & $13.8$ & $17.4$ & $2.3$ \\
$(m-M)_0$ & $15.11$ & $16.70$ & $15.95$ & $15.70$ & $16.20$ & $11.80$ \\
$R_{GC}$ (kpc) & $2.7$ & $13.7$ & $11.9$ & $18.0$ & $21.2$ & $6.3$ \\
$M_V$ & -3.8 $\pm$ 0.8 & -6.45 & -5.62 $\pm$ 1\\
E(B-V) & $1.09$ & $0.45$ & $0.61$ & $0.86$ & $1.17$ & $1.36$ \\
age (Gyr) & $>10$ & $<10$ & 11 $\pm$1 & $\sim$1.5 & $\sim$1 & $\sim$3\\
$\overline{v_{\rm los}} ~ (\kms)$ & $90.9\pm0.4$ & $-266.4\pm0.2$ & $31.0\pm0.1$ & $159.0\pm0.3$ & $135.9\pm0.4$ & $-9.8\pm0.1$ \\
$\sigma_v~(\kms)$ & $1.9\pm0.3$ & $1.4\pm0.2$ & $0.4\pm0.3$ & $1.0\pm0.3$ & $1.4_{-0.3}^{+0.4}$ & $0.6\pm0.1$ \\
$\overline{{\rm [Fe/H]}}$ & $-1.83_{-0.04}^{+0.03}$ & $-1.84\pm0.02$ & $-0.30\pm0.03$ & $-0.50\pm0.05$ & $-0.34\pm0.06$ & $0.10\pm0.03$ \\
$\sigma_{\rm [Fe/H]}$ & $<0.16$ & $<0.10$ & $<0.14$ & $<0.16$ & $<0.14$ & $<0.22$ \\
$\overline{\mu_{\alpha \star}}~(\masyr)$ & $-3.74\pm0.03$ & $0.51\pm0.01$ & $-4.35\pm0.02$ & $-1.08\pm0.03$ & $-0.73\pm0.03$ & $2.93_{-0.02}^{+0.01}$ \\
$\overline{\mu_{\delta }}~(\masyr)$ & $0.71_{-0.02}^{+0.01}$ & $-3.51\pm0.01$ & $-1.09\pm0.02$ & $1.50\pm0.03$ & $1.60_{-0.03}^{+0.04}$ & $0.44\pm0.02$ \\
$\varpi~(\mas)$ & $0.12\pm0.01$ & $0.07\pm0.01$ & $0.09\pm0.01$ & $0.08\pm0.01$ & $0.10\pm0.02$ & $0.437\pm0.005$ \\
$N_{\rm v_{\rm los}},~N_{\rm [Fe/H]}~N_{\rm \mu}$ & 35, 29, 33 & 62, 52, 65 & 43, 34, 42 & 19, 19, 19 & 23, 16, 21 & 80, 79, 86 \\
$r_{\rm peri}$ (kpc) & $2.9\pm1.0$ & $7.6_{-1.5}^{+1.6}$ & $9.8_{-1.8}^{+1.7}$ & $13.9_{-2.4}^{+2.7}$ & $19.9_{-2.5}^{+1.9}$ & $5.85\pm0.07$ \\
$r_{\rm apo}$ (kpc) & $3.3_{-0.4}^{+1.0}$ & $33.9_{-6.7}^{+8.8}$ & $13.3_{-1.4}^{+2.1}$ & $17.7\pm1.1$ & $23.9_{-3.4}^{+4.8}$ & $8.09\pm0.05$ \\
ecc & $0.07_{-0.02}^{+0.15}$ & $0.63_{-0.00}^{+0.01}$ & $0.16_{-0.02}^{+0.04}$ & $0.12\pm0.06$ & $0.10_{-0.01}^{+0.04}$ & $0.161\pm0.003$ \\
$P$ (Myr) & $-46_{-6}^{+7}$ & $-462_{-130}^{+97}$ & $-235_{-42}^{+35}$ & $-329_{-44}^{+39}$ & $-470_{-81}^{+68}$ & $-133\pm1$ \\
$z_{\rm max}$ (kpc) & $2.1_{-0.5}^{+0.9}$ & $20.5_{-3.7}^{+4.9}$ & $1.3\pm0.2$ & $1.4_{-0.2}^{+0.3}$ & $1.7_{-0.3}^{+0.4}$ & $0.18\pm0.01$ \\
${\rm E ~(10^{5}~km^{2}~s^{-2}})$ & $-1.77_{-0.13}^{+0.15}$ & $-0.84\pm0.09$ & $-1.12\pm0.07$ & $-0.99\pm0.05$ & $-0.85\pm0.06$ & $-1.371\pm0.004$ \\
${\rm L_Z ~(10^3~kpc~km~s^{-1}})$ & $0.47_{-0.13}^{+0.14}$ & $-0.49_{-0.39}^{+0.33}$ & $-2.51_{-0.38}^{+0.35}$ & $-3.39_{-0.41}^{+0.39}$ & $-4.60_{-0.59}^{+0.55}$ & $-1.59\pm0.01$ \\
		\hline
	\end{tabular}
\end{table*}

\begin{figure*}
\includegraphics[width=\textwidth]{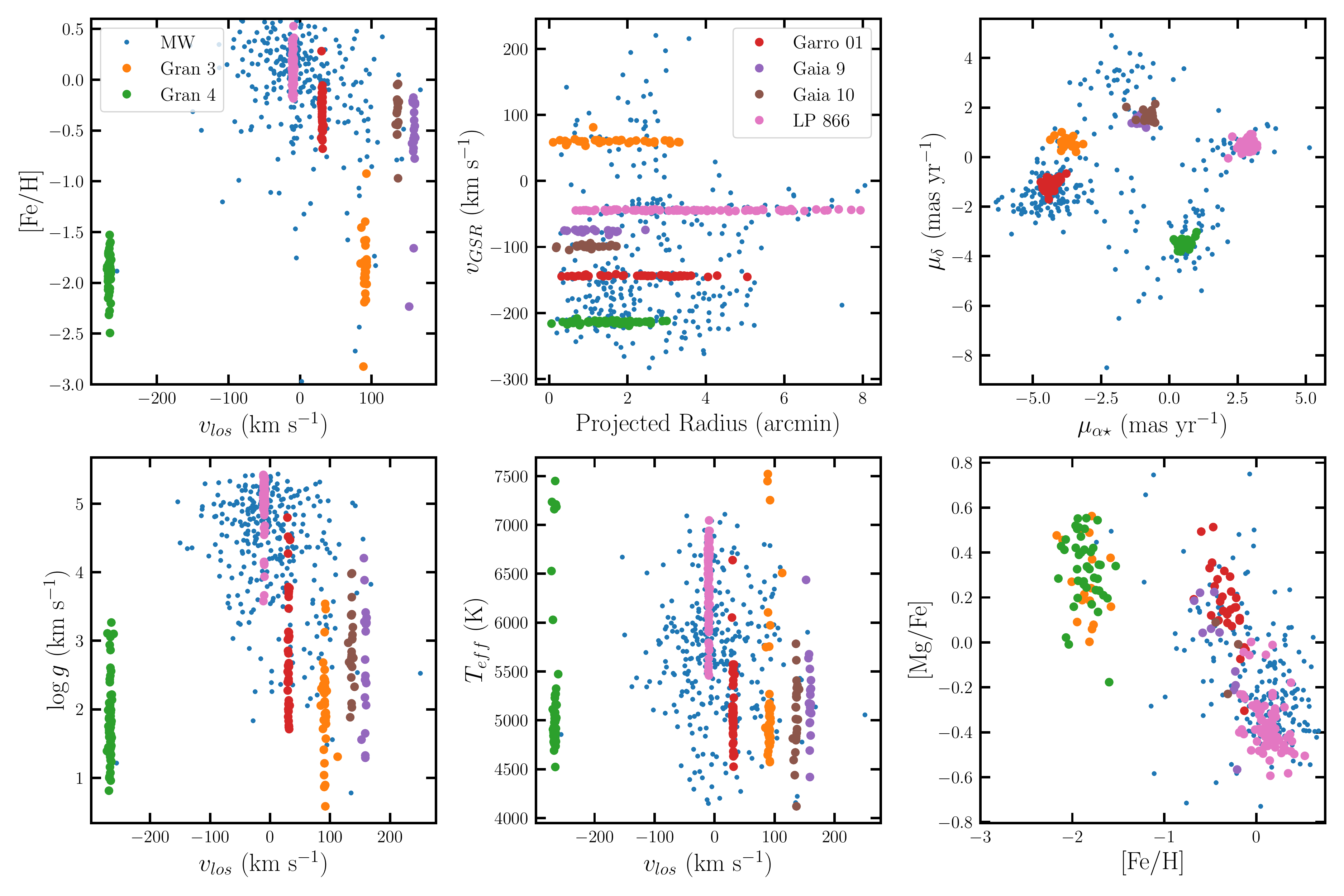}
\caption{Summary of the spectroscopic observations of the six star clusters and with members coloured for each star cluster. MW stars are blue points. The star clusters are clearly identified based on the narrow  $v_{ los}$ peaks. 
{\bf Top left panel:} line-of-sight velocity ($v_{los}$) versus metallicity ([Fe/H]). Only stars with good quality metallicity are included. 
{\bf Top middle panel:} Projected radial distance from the center for the cluster  versus velocity in the Galactic center of rest ($v_{GSR}$).
{\bf Top right panel:} vector point diagram ($\mu_{\alpha\star}$ vs $\mu_{\delta}$).
{\bf Bottom left panel:} $v_{los}$ versus surface gravity ($\log{ g}$).
{\bf Bottom middle panel:} $v_{los}$ versus effective temperature ($T_{\rm eff}$).
{\bf Bottom right panel:} [Fe/H] versus [Mg/Fe]. Only stars with good quality ${\rm [Fe/H]}$ and $[{\rm Mg}/{\rm Fe}]$ are included. 
}
\label{fig:kgo_summary}
\end{figure*}

In Figure~\ref{fig:kgo_summary}, we summarize the kinematics, chemistry, and stellar parameters of cluster members and MW foreground stars in our follow-up spectroscopic observations.  
The top panels from left to right compare the line-of-sight velocities ($v_{los}$) to the stellar metallicity ([Fe/H]), the radial distance from the center of the cluster versus the $v_{los}$, and the proper motion ($\mu_{\alpha\star}$, $\mu_{\delta}$).  The member stars in each cluster are highlighted in different  colours. Note in the proper motion panel, only stars with good quality astrometry are included. 
In the bottom  panels we compare $v_{los}$ to the surface gravity (left, $\log{g}$),  $v_{los}$ to effective temperature ($T_{\rm eff}$, center), and compare [Mg/Fe] vs [Fe/H].  Excluding \kgoeight, all stars are red giant branch/red clump stars (with several  horizontal branch stars in \kgotwo and \kgofour).  
The derived properties of the star clusters are summarized in Table~\ref{tab:kgo_properties}.

\subsection{Cluster Properties and Spectroscopic Membership}
\label{section:membership_properties}

We are able to identify the members of four clusters,  \kgotwo, \kgofour, \kgoseven, and \kgoten, based purely on the line-of-sight velocity as the cluster mean velocity is  distinct from the MW foreground.
The stellar parameters ([Fe/H], $T_{\rm eff}$, and $\log{g}$) and  photometry ($G$, $G_{BP}$, and $G_{RP}$) of the members identified from the velocities further reinforce their membership and they are  consistent with  single stellar populations.
While there is a clear overdensity in the $v_{\rm los}$ distribution in the \kgoeight and \kgotwotwo fields, there is overlap with the MW foreground   and we construct mixture models to quantitatively identify members in these objects.

\subsubsection{Gran~3/Patchick~125}
\label{sec:kgo2}

\begin{figure*}
\includegraphics[width=\textwidth]{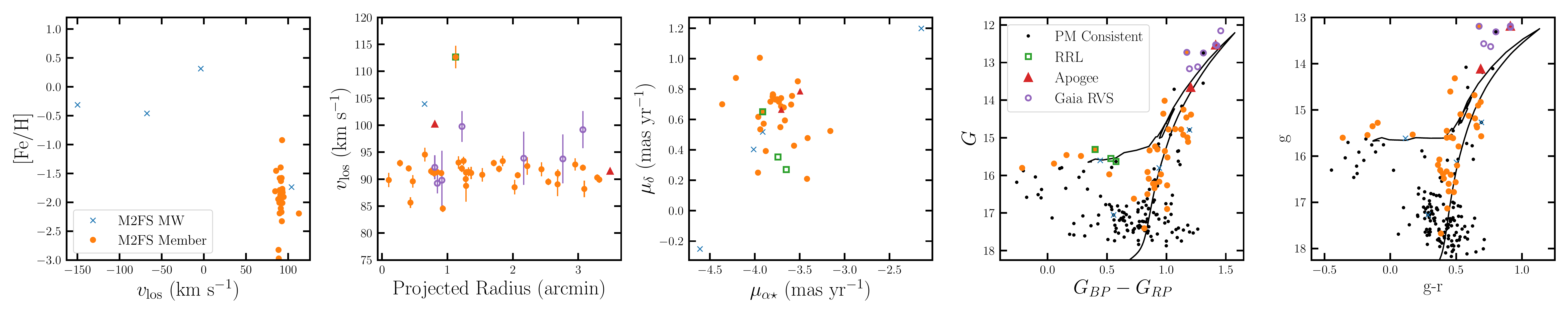}
\caption{Summary of the M2FS observations of Gran~3/Patchick~125. {\bf Left}: line-of-sight velocity ($v_{\rm los}$) versus metallicity ([Fe/H]) for  Gran~3 M2FS members (orange circles) and MW foreground stars (blue x's).   {\bf Center-left}: projected radius versus $v_{\rm los}$.  We denote the M2FS observation of a RRL with a green square.  The red triangles and purple circles are APOGEE and {\it Gaia} RVS members, respectively.  {\bf Center}:  Vector point diagram for the same stars. We include two additional RRL candidate members without spectroscopy.  {\bf Center-right}: {\it Gaia} colour magnitude diagram ($G_{BP}-G_{RP}$ versus $G$).  We  include all stars within 6\arcmin\ that are consistent with the proper motion and parallax of Gran~3 (small black points).  We include an isochrone with age = 13~Gyr and [Fe/H] = $-1.9$. For the photometry to match the isochrone, we increased the standard MW reddening law to  $R_V=3.3$. {\bf Right: } DECam g-r vs g photometry from DECaPS. 
}
\label{fig:kgo2_summary}
\end{figure*}

In Figure~\ref{fig:kgo2_summary}, we  summarize the \kgotwo members and the MW foreground stars from our M2FS observations of the \kgotwo field. 
In addition to the primary M2FS sample, we include 3 RRL members from the {\it Gaia} DR3 catalog \citep{Clementini2022arXiv220606278C},   2 APOGEE  members\footnote{First identified in \citet{FernandezTrincado2022A&A...657A..84F}.}, and 7 {\it Gaia } DR3 RVS members\footnote{6 of the 7 of the members were first identified in \citet{Garro2023A&A...669A.136G}.}.
The 36 M2FS members of \kgotwo are identified by selecting stars in the $ 82<v_{\rm los} < 97\kms$  velocity range. 
The M2FS stars all are consistent with a single metallicity  and the  stellar parameters ($\log{\rm g}$ and $T_{\rm eff}$) are consistent with red giant stars or horizontal branch stars. 
There are two stars outside this range within $\sim20~\kms$ of the mean velocity of \kgotwo. 
The first, source\_id\footnote{Here and throughout the paper source\_id refers to {\it Gaia} DR3 source\_id.}=5977223144516980608, is a RRL star and the distance of this star agrees with the star cluster. We consider it a cluster member but we  exclude this star from all kinematic analysis due to the  velocity variability of RRL stars. 
The second, source\_id=5977223144516066944, is a $7\sigma$ outlier in velocity but the stellar parameters agree with the cluster mean metallicity and the proper motion agrees with the cluster. We exclude it from our analysis and suggest that if it is a member, it is likely a binary star \citep[e.g.,][]{Spencer2018AJ....156..257S}.  Additional multi-epoch data is required to confirm this.

To determine the kinematics and chemistry we use a two-parameter Gaussian likelihood function \citep{Walker2006AJ....131.2114W} and to use \texttt{emcee} to sample from the posterior \citep{ForemanMackey2013PASP..125..306F}. We use a uniform prior for the average  and a Jeffreys prior for the  dispersion. 
From the 35 stars in the M2FS sample that are non-variable, we measure $\overline{v_{\rm los}} = +90.9\pm0.4 ~\kms$,  $\sigma_v=1.9\pm0.3~\kms$.
With the 29 members with good quality [Fe/H] measurements, we measure $\overline{{\rm [Fe/H]}}=-1.83_{-0.04}^{+0.04}$ and $\sigma_{\rm [Fe/H]} = 0.09\pm0.4$ ($\sigma_{\rm [Fe/H]}<0.16$). For  limits throughout this work, we list values at 95\% confidence intervals.
We note that the non-zero metallicity dispersion is due to one star (source\_id=5977224587625168768; $[{\rm Fe/H}]=-1.58\pm0.07$) that is $3.5\sigma$ larger than the mean metallicity of \kgotwo. This star has stellar parameters and mean velocity that are otherwise consistent with \kgotwo. If this star is removed, the kinematics and mean metallicity are unchanged but the metallicity dispersion is  constrained to less than  $\sigma_{\rm [Fe/H]}<0.10$. We have opted to include this star in our sample. 
From the 33 spectroscopically (M2FS, APOGEE, and {\it Gaia} RVS) identified members with good quality astrometry, we measure: $\overline{\mu_{\alpha \star}} =  -3.74\pm0.03\masyr$, 
$\overline{\mu_{\delta}} = 0.71_{-0.02}^{+0.01} \masyr$, $\sigma_{\mu_{\alpha \star}} = 0.10_{-0.02}^{+0.03}\masyr$, $\sigma_{\mu_{\delta}} = 0.03_{-0.01}^{+0.02}\masyr$, and $\varpi=0.12\pm0.01 \mas$. 
The parallax measurement corresponds  to $d=8.6_{-0.8}^{+0.9}~\kpc$ and $(m-M)_0=14.7\pm0.2$, which is closer than the isochrone or RRL distance (see below).
Assuming a distance of 10.5~kpc (in agreement with these latter measurements), the proper motion dispersions correspond to $\sigma_{\mu_{\alpha \star}}=5.0_{-1.1}^{+1.4}~\kms$ and $\sigma_{\mu_{\delta}}=1.7_{-0.7}^{+0.9}~\kms$.  While $\sigma_{\mu_{\alpha \star}}$ is larger than expected, $\sigma_{\mu_{\delta}}$ agrees with $\sigma_{v}$. 
With the 7 {\it Gaia} RVS members we measure: $\overline{v_{\rm los}} = +93.5_{-1.5}^{+1.7} ~\kms$ and  $\sigma_v<6.6~\kms$ (95\% c.i.). The {\it Gaia} RVS sample is consistent with the M2FS sample.

There are 3 stars in the secondary samples (1 APOGEE, 2 {\it Gaia} RVS) that are $\sim9~\kms$ offset ($1-5\sigma$ outliers in $v_{\rm los}$) from the bulk of \kgotwo. 
2 of these stars have repeat measurements with  other samples and those measurements are in good agreement  with the bulk velocity of the system and suggests that those stars could be binary stars. 
The APOGEE star (source\_id=5977223316333009024; $v_{\rm los,~APOGEE}=100.2\pm0.2\kms$) overlaps with {\it Gaia} RVS ($v_{\rm los,~{\it Gaia}~RVS}=92.2\pm2.2\kms$) and one of the {\it Gaia} RVS  members source\_id=5977224587625168768; ($v_{\rm los,~{\it Gaia}~RVS}=99.2\pm3.5\kms$) overlaps with M2FS ($v_{\rm los,~M2FS}=92.6\pm0.7\kms$). Both these stars may be binary stars which would explain their offset. 

We identify three RRL in  the {\it Gaia} DR3 RRL catalog (source\_id=5977223144516980608, 5977224553266268928\footnote{We note that this star is a 3-$\sigma$ outlier in $\mu_\delta$ compared to the systemic proper motion, however, the large value of \texttt{astrometric\_excess\_noise\_sig}  suggests that the astrometric solution may not be reliable.}, 5977224557581335424) that are consistent with the proper motion and spatial position (all three have $R<2\arcmin$) of \kgotwo.  One star (source\_id=5977223144516980608) was observed with M2FS.  It is offset from the mean  velocity of \kgotwo by $\sim20\kms$. As RRL stars are variable in velocity and vary more than $50\kms$  over the period \citep{Layden1994AJ....108.1016L}, we consider this star a member.  With more spectroscopic epochs the systemic velocity of the star could be measured \citep[e.g.,][]{Vivas2005AJ....129..189V}.  
We apply the metallicity correction to the absolute magnitude of a RRL in {\it Gaia} bands to determine the absolute magnitude:  $M_G = 0.32 {\rm [Fe/H] }+ 1.11$ \citep{Muraveva2018MNRAS.481.1195M}.
From the three RRL,  we find a mean distance modulus of $(m-M)_0=15.1$ corresponding to a distance of $d=10.5 \kpc$.  
This is slightly smaller than other distance measurements for this cluster: $d=12.02~\kpc$  \citep{Gran2022MNRAS.509.4962G}, $d=11\pm0.5~\kpc$ \citep{FernandezTrincado2022A&A...657A..84F}, and  $d=10.9\pm0.5,~11.2\pm0.5~\kpc$ \citep{Garro2022A&A...659A.155G}.

In Figure~\ref{fig:kgo2_summary}, we compare an old (age $=13$ Gyr) and metal-poor isochrone ([Fe/H] = $-1.9$) to \kgotwo with both {\it Gaia} and DECaPS photometry.
We are able to  match the horizontal branch and the colour of the RGB if we use $(m-M)_0 = 15.2$ and a  $R_V=3.3$ dust law (compared to a  standard of $R_V=3.1$) and find it difficult to match the horizontal branch using the RRL distance.
As noted by \citet{Garro2022A&A...659A.155G}, some studies suggest a lower dust law is favored in the bulge regions \citep[e.g.,][]{Souza2021A&A...656A..78S, Saha2019ApJ...874...30S} which would disagree with the RGB of \kgotwo. 
This larger distance modulus is in better agreement with the literature distance measurements of \kgotwo  \citep{Gran2022MNRAS.509.4962G, FernandezTrincado2022A&A...657A..84F, Garro2022A&A...659A.155G}.
We note that the brighter stars are bluer than the isochrone but the bulk of the RGB matches the isochrone.

\subsubsection{Gran~4}

\begin{figure*}
\includegraphics[width=\textwidth]{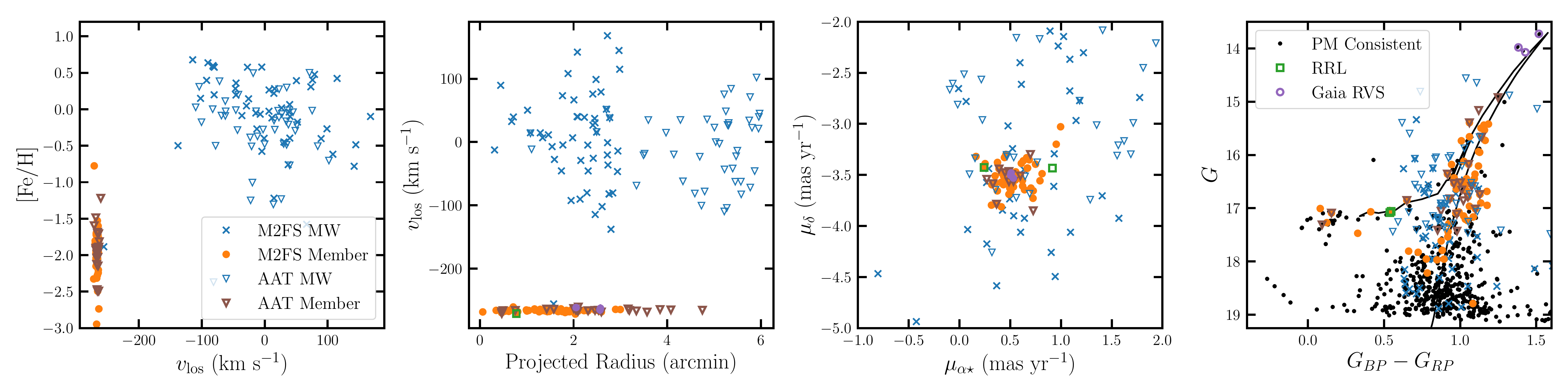}
\caption{Similar  to Figure~\ref{fig:kgo2_summary} but for \kgofour. We include AAT spectroscopic observations with brown and blue open triangles corresponding to Gran~4 members and MW stars, respectively. 
An isochrone with age = 13~Gyr and [Fe/H] = $-1.9$ is included as a black curve.
}
\label{fig:kgo4_summary}
\end{figure*}

The summary of our spectroscopic observations of \kgofour is shown in Figure~\ref{fig:kgo4_summary}. 
We identify 64, 22, and 3 \kgofour members in the M2FS, AAT and  {\it Gaia} RVS sample, respectively, with  the $ -280 < v_{\rm los} < -260\kms$ selection. 12  stars in the AAT sample overlap with the M2FS sample.

Within the M2FS sample, further examination of the velocity distribution reveals two  outlier stars.  
The first (source\_id=4077796986282497664) is a RRL star and is offset from the mean velocity by $\sim 8~\kms$ and by $\sim3\sigma$  once the velocity dispersion is considered. 
It is a cluster member but due to its variable nature, we exclude it from any kinematic analysis. The second (source\_id=4077796810168905344) is a $\sim 8 \sigma$ outlier in velocity and if it is considered a member the velocity dispersion increases from $1.2~\kms$ to $2~\kms$.  It is consistent with the mean metallicity and proper motion of \kgofour, and the stellar parameters ($T_{\rm eff}$  and $\log{g}$) are consistent with a red giant branch star.  It seems unlikely for this star to be MW star based on the MW velocity and metallicity distribution, however, as it is a $\sim 8 \sigma$ outlier it is either a binary star or a MW interloper and we exclude it from the analysis.

From the 62  non-variable M2FS members we find,  $\overline{v_{\rm los}} = -266.4\pm0.2 ~\kms$,  $\sigma_v=1.4\pm0.2~\kms$, $\overline{{\rm [Fe/H]}} = -1.84\pm0.02$ and $\sigma_{\rm [Fe/H]} < 0.10$.
Due to tail to zero dispersion and a lack of a clear peak, we do not consider the metallicity dispersion resolved and list an upper limit.
The stellar parameters of these stars ($T_{\rm eff}$ and $\log{g}$) are consistent with red giant branch stars or horizontal branch stars which further confirms our membership identification.  
With the 3 {\it Gaia} RVS members we measure: $\overline{v_{\rm los}} = -262.7_{-3.7}^{+3.6} ~\kms$,  $\sigma_v<6.2~\kms$ (95\% c.i.).

From the 22 AAT members, we find $\overline{v_{\rm los}} = -265.9 \pm 0.4 ~\kms$ and $\sigma_v=1.5_{-0.3}^{+0.4}~\kms$. 12 of these stars  overlap with the M2FS sample.  There is one velocity outlier (source\_id=4077796397852026240) with $\sim3\sigma$, however, this star is in the M2FS sample with almost the exact same velocity. Removing this star decreases the velocity dispersion to $\sim1~\kms$ from $\sim1.5~\kms$ in the AAT sample but  its inclusion or exclusion does not affect the M2FS kinematics and we opt to include it. 
From the calcium triplet metallicities, we compute $\overline{{\rm [Fe/H]}} = -1.82\pm0.06$ and $\sigma_{\rm [Fe/H]} =0.21_{-0.05}^{+0.06}$. The metallicity dispersion is clearly resolved in contrast to our expectation of a single stellar population in a star cluster and the lack of a metallicity dispersion in the M2FS sample. 
The source of the metallicity dispersion in the AAT data is unclear. 
As the M2FS sample is larger, we adopt the velocity and metallicity results from the M2FS sample as our primary results. 

From the combined M2FS,  AAT, and {\it Gaia} RVS sample there are 65 stars with good quality astrometric measurements. 
From these stars we measure:  $\overline{\mu_{\alpha \star}} = +0.51\pm0.01 \masyr$, and
$\overline{\mu_{\delta}} = -3.51\pm0.01 \masyr$, $\sigma_{\mu_{\alpha \star}} = 0.02_{-0.01}^{+0.03}\masyr$, $\sigma_{\mu_{\delta}} =0.05_{-0.03}^{+0.02} \masyr$, and $\varpi=0.07\pm0.01 \mas$.
The parallax measurement corresponds  to $d=14.6_{-1.9}^{+2.5}~\kpc$ and $(m-M)_0=15.8\pm0.3$ which is closer than the other distance measurements (see below).
Assuming a distance of 21.9~kpc (from the best-fit isochrone distance), the proper motion dispersion correspond to: $\sigma_{\mu_{\alpha \star}} = 2.6_{-1.2}^{+2.7}\kms$ and $\sigma_{\mu_{\delta}} =5.4_{-2.8}^{+2.5} \kms$.

We identify two RRL (source\_id=4077796986282497664, 4077796573965756928) in the {\it Gaia} DR3 RRL catalog \citep{Clementini2022arXiv220606278C} as members of \kgofour based on their proper motion and distance. 
The first RRL is also in the VVV RRL \citep{Molnar2022MNRAS.509.2566M} and  OGLE RRL catalogs \citep{Soszynski2019AcA....69..321S}.
There is a third candidate RRL (source\_id=4077796608325479168) from the PanSTARRS1 RRL catalog \citep{Sesar2017AJ....153..204S}, however, it is not in the {\it Gaia} DR3 RRL catalog so we do not include it in the analysis. It is considered a variable star in  {\it Gaia} DR3 but only has a \texttt{best\_class\_score}=0.4 for being an RRL. 
Additional time series data are required to confirm the status of this star. 
From the two RRL members, $\mu=16.5, 16.6$ corresponding to $d=19.9, 21~\kpc$. This is slightly closer than our best-fit  $\mu\sim16.7$ from matching the horizontal branch to a metal-poor isochrone. It is also closer than $\mu=16.84$, $d=22.49~\kpc$ from \citet{Gran2022MNRAS.509.4962G}.

\subsubsection{Garro~01}

\begin{figure*}
\includegraphics[width=\textwidth]{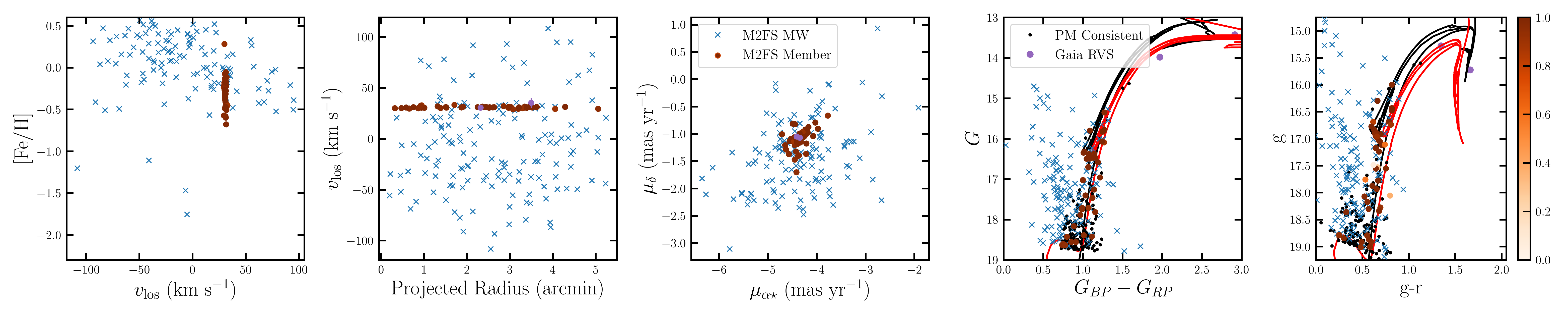}
\caption{
Same as Figure~\ref{fig:kgo2_summary} but for \kgotwotwo. In contrast to the other globular clusters (\kgotwo and \kgofour) we use a mixture model to identify \kgotwotwo members and stars with membership $>0.01$ are coloured according to their membership probability in the colour bars in the center-right and right panels.
We include an isochrone from the photometric analysis of \citet{Garro2020A&A...642L..19G} with an age = 11 Gyr and [Fe/H] = $-0.7$ (black)  and an isochrone with our best estimate using the spectroscopic metallicity ([Fe/H]= -0.3) and an age = 4 Gyr (red).  There are no candidate RRL stars in \kgotwotwo.
}
\label{fig:kgo22_summary}
\end{figure*}

There is  overlap  in the MW foreground and star cluster velocity distributions for \kgotwotwo and \kgoeight and  we construct mixture models to account for this overlap.
The total likelihood for our mixture models is:

\begin{equation}
\mathcal{L} =  f_{\rm cluster} \mathcal{L}_{\rm cluster} + (1 - f_{\rm cluster}) \mathcal{L}_{\rm MW} \, ,
\label{eqn:mixture}
\end{equation}

\noindent where $\mathcal{L}_{\rm cluster} $, $\mathcal{L}_{\rm MW}$, and $f_{\rm cluster}$ correspond to the cluster population, the MW population, and the fraction of stars in the star cluster, respectively \citep[e.g.,][]{Pace2020MNRAS.495.3022P, Pace2021ApJ...923...77P}.   
We assume the  probability distributions of each data component are separable:

\begin{equation}
\mathcal{L}_{\rm cluster/MW} =  \mathcal{L}_{\rm spatial} \mathcal{L}_{\rm PM} \mathcal{L}_{\rm v_{los}} \mathcal{L}_{\rm [Fe/H]}  \, . 
\label{eqn:likelihood}
\end{equation}

\noindent Where $\mathcal{L}_{\rm spatial}$, $\mathcal{L}_{\rm PM}$, $\mathcal{L}_{\rm v_{los}}$, and $\mathcal{L}_{\rm [Fe/H]}$ are the spatial likelihood,  the proper motion likelihood, the line-of-sight velocity likelihood, and the metallicity likelihood, respectively.
To compute the membership of each star, we compare the ratio of the cluster likelihood to total likelihood for each star: $p_{\rm member} = \mathcal{L}_{\rm cluster}/\mathcal{L}$ \citep[e.g.,][]{Martinez2011ApJ...738...55M}.

For \kgotwotwo, we primarily analyze a mixture model with $v_{\rm los}$  and [Fe/H] but also consider a second model with spatial information. 
We assume the $v_{\rm los}$  and [Fe/H] likelihood distributions  are Gaussian for both the star cluster and MW components.
We  apply the following cuts to the \kgotwotwo spectroscopic sample to remove MW foreground stars: $T_{\rm eff}-2\times \sigma_{T_{\rm eff}}<6000 K$, $\log_{10}{g}-2\times \sigma_{\log_{10}{g}}<4$, a parallax cut ($\varpi - 3 \sigma_{\varpi}< 0.064$), and  a loose  $G_{BP}-G_{RP}$ colour cut of  0.25  around an age = 11 Gyr and [Fe/H] = $-0.6$ MIST isochrone \citep{Dotter2016ApJS..222....8D}. 
The isochrone selection is applied to remove  blue MW main sequence stars from the sample which have no overlap in colour-magnitude space with the cluster members. 
From our primary mixture model, we find: $\overline{v_{\rm los}} = +31.0\pm0.1~\kms$,  $\sigma_v=0.4\pm0.3~\kms$ ($\sigma_v < 0.8~\kms$), $\overline{{\rm [Fe/H]}}=-0.30\pm0.03$ and $\sigma_{\rm [Fe/H]}<0.14$.
In Figure~\ref{fig:kgo22_summary}, we summarize the properties of the spectroscopic members identified in the mixture model. Stars are coloured by their membership probability.

For the second spatial model we use a conditional likelihood and we assume that the fraction of stars is spatially dependent  \citep[e.g.,][]{Martinez2011ApJ...738...55M, Pace2021ApJ...923...77P}:
\begin{equation}
f_{\rm cluster}(R) = \Sigma_{\rm cluster}(R)/(\Sigma_{\rm cluster}(R) + N \Sigma_{\rm MW}(R)). \\ 
\end{equation}
Where $\Sigma$ is the projected stellar distribution and $N$ is the relative normalization between the cluster and MW spatial distributions
We assume a \citet{King1962AJ.....67..471K} distribution for the \kgotwotwo distribution with parameters from \citet{Garro2020A&A...642L..19G} and assume that the MW distribution is constant over the small region examined. We utilize a  conditional likelihood as there is an unknown spatial selection function. 

While the  chemodynamic properties of \kgotwotwo from the two models are nearly identical, there are some minor differences in the  membership of individual stars between the two models.
The conditional likelihood model has larger membership for stars near the center and lower membership for the two most distant stars. 
The difference in membership between the models for individual stars is small ($\sum \vert p_{\rm standard} - p_{\rm conditional}\vert = 1.7$) and the overall membership is similar for the two models; both models  have $\sum p = 46.8\pm1.9$. 
We consider stars with $p>0.9$ as high confidence members and there are 39 and 42  members identified in the standard model and  conditional likelihood model, respectively and  note there are  43  high confidence members with $p>0.9$ in either  model.

From the 42 members with good astrometry (42 M2FS and 1 {\it Gaia} RVS), we measure: $\overline{\mu_{\alpha \star}} = -4.35\pm0.02~\masyr$, 
$\overline{\mu_{\delta}} = -1.09\pm0.02~\masyr$, $\sigma_{\mu_{\alpha \star}} = 0.09_{-0.02}^{+0.02}~\masyr$, and $\sigma_{\mu_{\delta}} = 0.08_{-0.02}^{+0.03}~\masyr$. 
We measure $\varpi=0.08\pm0.01 \mas$ corresponding to $d=11.9_{-1.5}^{+2.1}~\kpc$ and $(m-M)_0=15.4\pm0.3$.
The distance from the parallax is closer than the measurement derived from isochrone fits. 
Assuming a distance of 15.3~kpc from \citet{Garro2020A&A...642L..19G}, the proper motion dispersion terms correspond to: $\sigma_{\mu_{\alpha \star}} = 6.6_{-1.4}^{+1.5}~\kms$ and $\sigma_{\mu_{\delta}} = 5.8_{-1.8}^{+1.8}~\kms$. 
Our proper motion measurement  agrees with the {\it Gaia} DR2 proper motion measurement, $\overline{\mu_{\alpha \star}} = -4.68\pm0.47~\masyr$, and 
$\overline{\mu_{\delta}} = -1.35\pm0.45~\masyr$ from \citet{Garro2020A&A...642L..19G}.

We identify 2 members in the {\it Gaia} RVS sample with velocities and proper motion consistent with \kgotwotwo. Both stars are more evolved than the M2FS sample but roughly match the isochrone. Both stars are included in Figure~\ref{fig:kgo22_summary} as purple circles. 

We do not identify any RRL that have the same proper motion or a consistent distance with \kgotwotwo. 
With the distance modulus of \citet{Garro2020A&A...642L..19G}, we can match the red clump of our spectroscopic sample with an isochrone and we adopt this distance for our analysis. 

In Figure~\ref{fig:kgo22_summary}, we show optical colour-magnitude diagrams using {\it Gaia} and DECaPS  photometry. The prominent feature in the CMDs is the red clump and complements the near-infrared discovery photometry from \citet{Garro2020A&A...642L..19G}.  
Our spectroscopic metallicity  measurement is more metal-rich  than the isochrone analysis of \citet{Garro2020A&A...642L..19G} which found [Fe/H] = $-0.7$ with their CMDs fits (black isochrone in Figure~\ref{fig:kgo22_summary}). 
However, we cannot match the colour of the system with this age and spectroscopic metallicity. If we assume the spectroscopic metallicity, the isochrone is redder than the photometry. We discuss this in more detail in Section~\ref{sec:nature_kgo22}.

\subsubsection{Gaia~9}

\begin{figure*}
\includegraphics[width=\textwidth]{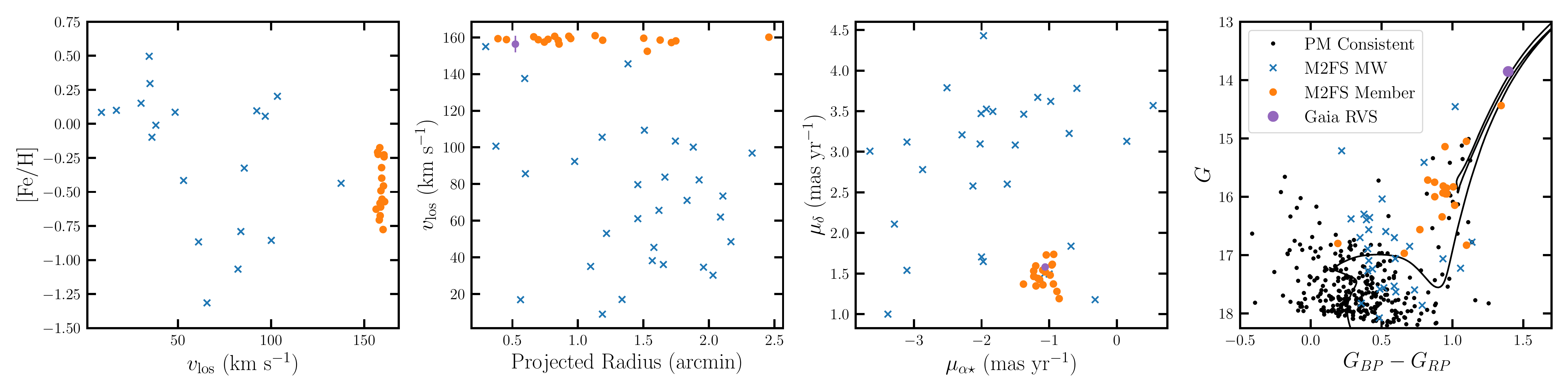}
\caption{Similar  to Figure~\ref{fig:kgo2_summary} but for \kgoseven.
The best fit isochrone is age = 1.5 Gyr and [Fe/H] = $-0.5$. 
}
\label{fig:kgo7_summary}
\end{figure*}

We identify 19 M2FS members of \kgoseven with the velocity selection: $ 150  < v_{\rm los} < 170~\kms$. 
There is one star inside this velocity range (source\_id=5537860050401680000) that is a non-member based on its proper motion. 
The properties of the members and proper motion selected stars are displayed in Figure~\ref{fig:kgo7_summary}.
From the 19  members, we measure:  $\overline{v_{\rm los}} = +159.0\pm0.3~\kms$,  $\sigma_v=1.0\pm0.3~\kms$, $\overline{{\rm [Fe/H]}}=-0.50\pm0.06$ and $\sigma_{\rm [Fe/H]}<0.16$.
There is one additional member {\it Gaia} RVS that we identify (source\_id=5537859848550503168) based on the radial velocity and proper motion and we include it in Figure~\ref{fig:kgo7_summary}.
From the 19  members with good astrometry, we measure: $\overline{\mu_{\alpha \star}} = -1.08\pm0.03~\masyr$, $\overline{\mu_{\delta}} = +1.50\pm0.03~\masyr$, $\sigma_{\mu_{\alpha \star}} = 0.07\pm0.04~\masyr$, $\sigma_{\mu_{\delta}} =0.08_{-0.03}^{+0.03}~\masyr$, and $\varpi=0.08\pm0.01 \mas$.
The parallax measurement corresponds to $d=12.9_{-2.0}^{+3.0}~\kpc$ and $(m-M)_0=15.6_{-0.4}^{+0.5}$ which agrees with our isochrone derived distance (see below).
Assuming a distance of 13.8~kpc (from our isochrone derived distance), the proper motion dispersion terms correspond to: $\sigma_{\mu_{\alpha \star}} = 4.8_{-2.4}^{+2.4}~\kms$ and $\sigma_{\mu_{\delta}} = 5.0_{-2.1}^{+2.3}~\kms$.

In the {\it Gaia} colour-magnitude diagram (Figure~\ref{fig:kgo7_summary}), the majority of the spectroscopic members are red clump stars.  We estimate the distance of the cluster to be $(m-M)_0=15.7$ or $d=13.8~\kpc$ based on the red clump using an MIST isochrone with age = 1.5~Gyr and [Fe/H] = $-0.5$.
The  blue stars at $G_0\sim17$ are likely the top of the main sequence and we use this feature to assist in estimating the age of the system. 
\kgoseven is younger than the other clusters examined thus far and it likely an open cluster.

\subsubsection{Gaia~10}

\begin{figure*}
\includegraphics[width=\textwidth]{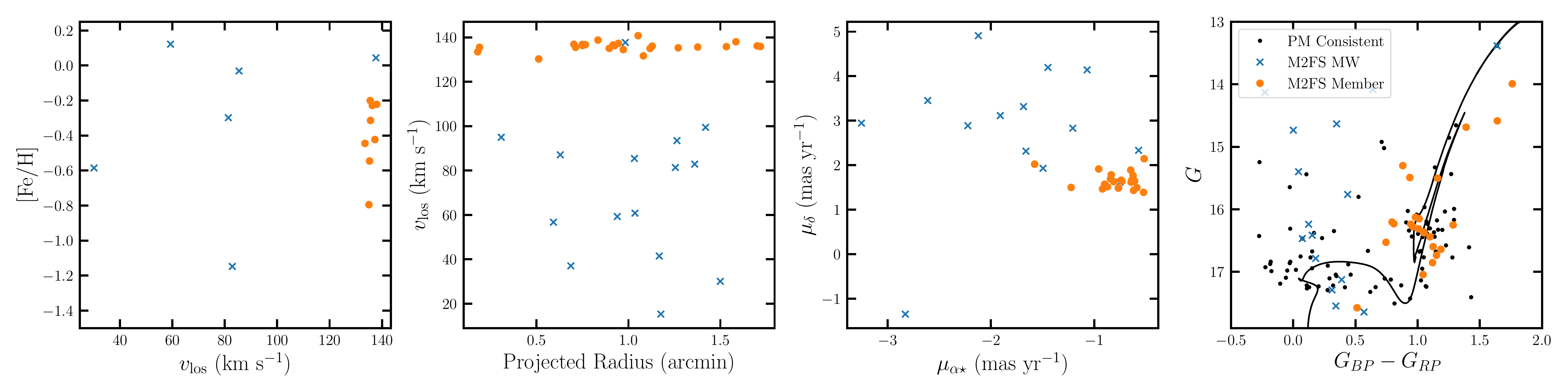}
\caption{Similar  to Figure~\ref{fig:kgo2_summary} but for \kgoten. 
The best fit isochrone is age = 1 Gyr and [Fe/H] = $-0.5$. Similar to \kgotwo, we  increase the extinction  to find an adequate  fit ($R_V=3.3$). 
}
\label{fig:kgo10_summary}
\end{figure*}

\kgoten has similar properties to \kgoseven but is  more distant. 
We summarize our spectroscopic sample in Figure~\ref{fig:kgo10_summary}.
We identify 23 members of \kgoten with a velocity selection: $ v_{\rm los}  > 120\kms$. 
Two stars (source\_id=5534905976200827776 and 5540909996882971520)  have kinematics (line-of-sight velocity and proper motion) that agree with the cluster but both stars are significant outerliers in metallicity. 5534905976200827776 is $3.2\sigma$ more metal-rich while 5540909996882971520 is $4.9\sigma$ more metal-poor and the inclusion of either star results in an offset mean metallicity and non-zero metallicity dispersion. We consider both stars non-members.
The  inclusion of the star as a members would infer to a non-zero  metallicity dispersion. We note that this same star is in the {\it Gaia} RVS catalog with a similar velocity. 
From the spectroscopic members, we measure:  $\overline{v_{\rm los}} = +135.9\pm0.4~\kms$,  $\sigma_v=1.4_{-0.3}^{+0.4}~\kms$, $\overline{{\rm [Fe/H]}}=-0.34 \pm0.06$ and $\sigma_{\rm [Fe/H]}<0.14$.

From the 21 members with good astrometry, we measure: $\overline{\mu_{\alpha \star}} = -0.74\pm0.03~\masyr$, 
$\overline{\mu_{\delta}} = +1.60_{-0.03}^{+0.04}~\masyr$, $\sigma_{\mu_{\alpha \star}} = 0.05_{-0.03}^{+0.05}~\masyr$, $\sigma_{\mu_{\delta}} =0.06_{-0.03}^{+0.04}~\masyr$, and $\varpi=0.09\pm0.02 \mas$.
The parallax measurement corresponds to $d=10.8_{-1.9}^{+3.1}~\kpc$ and $(m-M)_0=15.2_{-0.4}^{+0.6}$ which  is much closer than  our isochrone derived distance (see below).
Assuming a distance of 17.4~kpc (from our isochrone derived distance), the proper motion dispersion terms correspond to: $\sigma_{\mu_{\alpha \star}} = 4.2_{-2.6}^{+3.8}~\kms$ and $\sigma_{\mu_{\delta}} = 5.0_{-2.8}^{+3.5}~\kms$.

We compare theoretical isochrones based on the spectroscopic metallicity to the {\it Gaia} colour-magnitude diagram Figure~\ref{fig:kgo10_summary}. 
With an age of $1~{\rm Gyr}$ and distance modulus of $(m-M)_0=16.2$ ($d=17.4~\kpc$) we can fit the red clump and the possible main sequence turnoff based on the proper motion selected sample.  
Similar to \kgotwo, to match the colour of the isochrone a larger extinction coefficient of $R_V=3.3$ is required. There remains considerable spread in the colour of the members. This may be due to differential reddening.
We consider \kgoten an open cluster.

\subsubsection{LP~866}

\begin{figure*}
\includegraphics[width=\textwidth]{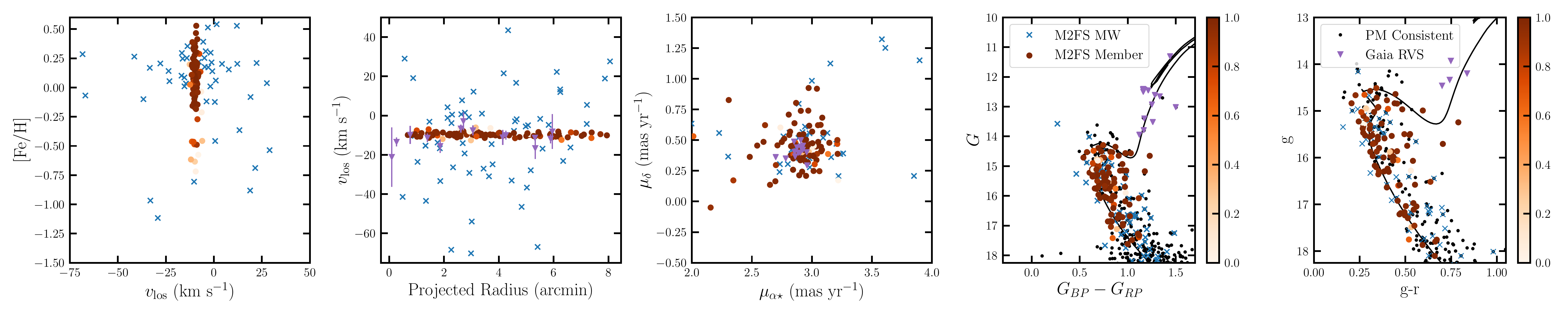}
\caption{
Same as Figure~\ref{fig:kgo22_summary} but for \kgoeight.  
Similar to \kgotwotwo, we include a colourbar for the membership from the mixture model.
The best fit isochrone is age = 3~Gyr and [Fe/H] = $+0.1$ but with a reduced extinction ($\sim 53\%$ of the E(B-V) value). We include DECam g-r vs g photometry from DECaPS. 
}
\label{fig:kgo8_summary}
\end{figure*}

\kgoeight is the only star cluster where the entirety of the M2FS spectroscopic sample is located on the main-sequence.
Similar to \kgotwotwo, we run a mixture model to account for the MW foreground distribution with $v_{\rm los}$,  proper motion,  and [Fe/H] components.
We do not include a conditional likelihood run as the spatial distribution was not known beforehand.
We assume the proper motion distribution is a truncated multi-variate Gaussian for both components with limits: $1.5 < \mu_{\alpha\star} < 3.5~{\rm mas~yr^{-1}}$, $-0.4 < \mu_{\delta} < 1.2~{\rm mas~yr^{-1}}$. While the proper motion was included in the spectroscopic target selection (based on {\it Gaia} DR2 astrometry), the proper motion dispersion is resolved in contrast to the other star clusters. 
We excluded stars outside of this proper motion limit and two bright stars with discrepant parallax measurements. 

With the mixture model, we identify 80 high confidence members  ($p>0.9$) and measure the following properties for \kgoeight: $\overline{v_{\rm los}} = -9.8\pm0.1~\kms$,  $\sigma_v=0.6\pm0.1~\kms$, $\overline{{\rm [Fe/H]}}=0.10\pm0.03$, $\sigma_{\rm [Fe/H]}=0.15\pm0.04$ ($\sigma_{\rm [Fe/H]}<0.22$), $\overline{\mu_{\alpha \star}} = 2.93_{-0.02}^{+0.01}~\masyr$, $\overline{\mu_{\delta}} = 0.44_\pm0.02~\masyr$,   $\sigma_{\mu_{\alpha \star}} =0.08_{-0.02}^{+0.02}~\masyr$, and $\sigma_{\mu_{\delta}} =0.10_{-0.01}^{+0.02}~\masyr$. 
The overall membership is $\sum p = 86.4\pm4.8$. As there is a non-zero metallicity dispersion, it is possible that our model has incorrectly identified some MW stars as cluster members.
From the 86 members with good astrometry, we measure $\varpi=0.437\pm0.005 \mas$ corresponding to $d=2.29\pm0.03~\kpc$ and $(m-M)_0=11.80\pm0.02$.
Assuming a distance of 2.29~kpc, the proper motion dispersion terms correspond to: $\sigma_{\mu_{\alpha \star}} = 0.9\pm0.02~\kms$ and $\sigma_{\mu_{\delta}} = 1.1\pm0.2~\kms$.

In the {\it Gaia} RVS sample there is a clear overdensity and velocity peak distinct from the MW population within  the central 8\arcmin of \kgoeight. 
We find   14 stars in the {\it Gaia} RVS catalog that are consistent with the velocity, proper motion, and parallax of \kgoeight.  All 14 stars are more evolved than the M2FS sample and aid in matching a stellar isochrone. 
With the 14 {\it Gaia} RVS members we find: $\overline{v_{\rm los}} = -10.6\pm0.8~\kms$, and  $\sigma_v=0.2_{-0.1}^{+0.9}~\kms$ ($\sigma_v < 2.2~\kms$); and from the 12 stars with good quality astrometry we find: $\overline{\mu_{\alpha \star}} = 2.88\pm0.01~\masyr$, and $\overline{\mu_{\delta}} = 0.44\pm0.02~\masyr$.  These results are consistent with the M2FS sample.

We include   {\it Gaia} and DECaPs colour magnitude diagrams in Figure~\ref{fig:kgo8_summary}. 
We match theoretical isochrones to estimate the age of the system. Unlike the other two open clusters, we have a confident distance and metallicity measurement.
We find that reducing the E(B-V) by $\sim53\%$  and setting the to age = $10^{9.5}$ = $\sim$~3~Gyr provides an adequate match. 
We note that the we varied the extinction to match the RGB of the {\it Gaia} RVS stars and varied the age to match the main sequence turn-off of the M2FS sample with the {\it Gaia} photometry. 
The same isochrone provided an similarly adequate match to the g-r vs g DECaPs photometry. 
More in depth modeling is required to improve constraints on the  age and extinction of the cluster. Regardless, we consider \kgoeight an open cluster.

\subsection{Spatial Distribution}

\begin{figure*}
\includegraphics[width=\textwidth]{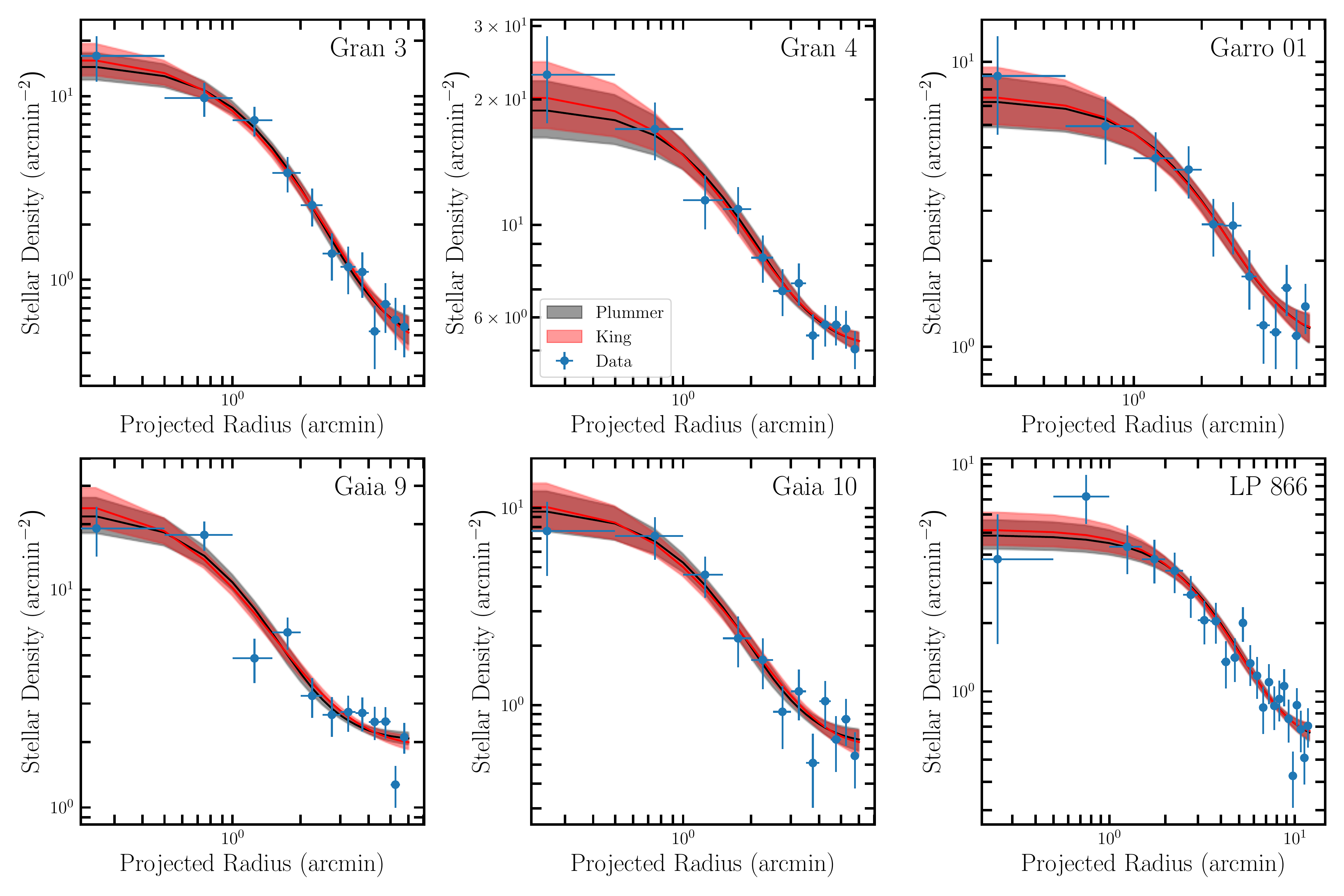}
\caption{ Projected radial stellar density profile of our  star cluster sample. 
{\bf Top:} the globular clusters, from left to right: \kgotwo, \kgofour, \kgotwotwo. 
{\bf Bottom:} the open clusters, from left to right: \kgoseven, \kgoten, \kgoeight. 
}
\label{fig:spatial_density}
\end{figure*}

To measure the spatial distribution of each star cluster, we construct a larger proper motion selected sample from the {\it Gaia} DR3 catalog based on the systemic proper motion   found from our spectroscopic sample and  apply a spatial mixture model (Equations~\ref{eqn:mixture}, \ref{eqn:likelihood}). 
We do not apply this methodology to the spectroscopic sample as it is not spatially complete and has an unknown spatial target selection. 
For the spatial likelihood we model the star cluster with two density profiles. The first profile is a Plummer distribution  \citep{Plummer1911MNRAS..71..460P}:

 \begin{equation}
    \Sigma(R) = \frac{1}{\pi r_p^2} \frac{1}{\left(1 + \left(R/r_p\right)^2 \right)^{2}}
\end{equation}

\noindent where $r_p$ is the Plummer scale radius (for a Plummer profile $r_p$ is equivalent to the 2D deprojected half-light radius).
The second is the King profile \citep{King1962AJ.....67..471K}:

 \begin{equation}
	\Sigma_{\star} (R) \propto \left[ \left(1 + \frac{R^2}{r_c^2} \right)^{-1/2} - \left(1 + \frac{r_t^2}{r_c^2} \right)^{-1/2} \: \right]^2
 \end{equation}

\noindent where  $r_c$ is the core radius and $r_t$ is the tidal radius.
We model a small region near each cluster and assume that the MW background is  constant within that small area after a proper motion selection is applied.

For the {\it Gaia} selected sample, we apply the following cuts: a 3-$\sigma$ selection in proper motion,   a parallax selection  ($\varpi - \varpi_{\rm cluster} - 3 \sigma_{\varpi} < 0$), $G<20$, $R<R_{\rm max}$, and stars with  good astrometry (i.e., satisfy our astrometric cuts in Section~\ref{section:gaia}). 
We will refer to this {\it Gaia} selected  sample and utilize the same sample for examining the colour-magnitude diagrams of the clusters. 
We use $R_{\rm max}=12\arcmin$ for \kgoeight and $R_{\rm max}=6\arcmin$ for all other clusters. 
For \kgotwotwo we additionally apply a loose  $G_{BP}-G_{RP}$ colour  cut of  0.25  around an age = 11 Gyr and [Fe/H] = $-0.6$ MIST isochrone \citep{Dotter2016ApJS..222....8D} following the spectroscopic selection. For the other clusters, the above selection primarily identifies stars with a  stellar population that agrees with the spectroscopic sample. 
Any photometric outliers (i.e., MW stars) will be roughly distributed uniformly within the small area examined and not bias the spatial distribution calculations. 
  
The Plummer and King fits along with the binned stellar profile of all six clusters are shown in Figure~\ref{fig:spatial_density}.  
In general, the results from  the Plummer and King  profile fits agree and provide adequate fits. Due to the low number of stars, there is no preference for one profile over the other.   We are unable to constrain $r_t$ and generally only provide lower limits. 
For the globular clusters, we find $r_h=1.7\pm0.2\arcmin$, $r_h=2.2_{-0.4}^{+0.5}\arcmin$, and $r_h = 2.4_{-0.4}^{+0.6}~\arcmin$ corresponding to $r_h=5.3_{-0.6}^{+0.7}~\pc$, $r_h=14.2_{-2.5}^{+3.3}~\pc$, and $r_h = 10.9_{-2.0}^{+2.6}~\pc$ from the Plummer profile fits for Gran~3, Gran~4, and \kgotwotwo, respectively.
With the King profile, we find $r_c=1.1_{-0.2}^{+0.3}\arcmin$, $r_c=1.4_{-0.4}^{+0.5}\arcmin$, and $r_c=1.8_{-0.5}^{+0.7}~\arcmin$ for Gran~3, Gran~4, and \kgotwotwo, respectively.
For comparison, \citet{Gran2022MNRAS.509.4962G} find $r_h=1.05\pm0.04\arcmin$ and  $r_h=1.14\pm0.02\arcmin$, for  Gran~3 and  Gran~4, respectively. These are smaller than the sizes we infer. 
\citet{Garro2020A&A...642L..19G} measure $r_c=2.5\pm1.5\arcmin$ for \kgotwotwo and a poorly constrained $r_t$ which agrees with our measurement. 

For the  open clusters, we find $r_h=1.4\pm0.2\arcmin$, $r_h=1.6_{-0.2}^{+0.3}\arcmin$, and $r_h=4.6_{-0.6}^{+0.7}\arcmin$ with the Plummer profile fits for \kgoseven, \kgoten, and \kgoeight, respectively. 
With the King profile, we find $r_c=0.8\pm0.2\arcmin$, $r_c=1.0_{-0.2}^{+0.3}\arcmin$, and $r_c=3.3_{-0.7}^{+0.8}~\arcmin$ for \kgoseven, \kgoten, and \kgoeight, respectively. 
The results for the six clusters are included in Table~\ref{tab:kgo_properties}.

\subsection{Orbital Properties}

\begin{figure*}
\includegraphics[width=\textwidth]{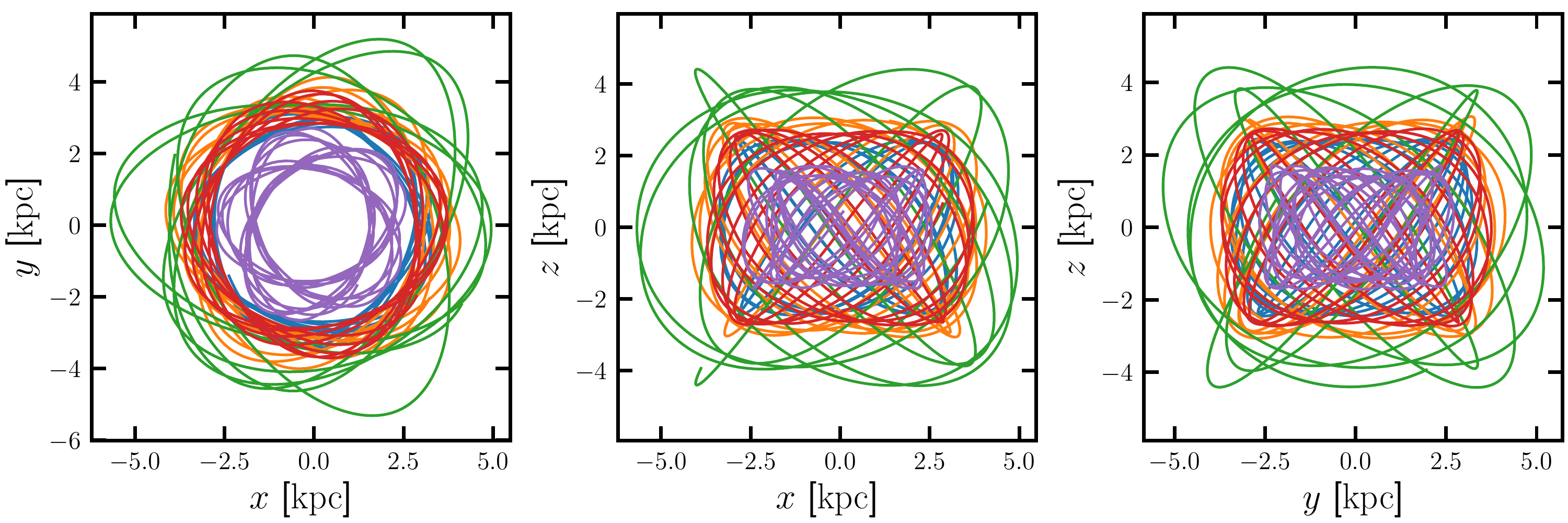}
\includegraphics[width=\textwidth]{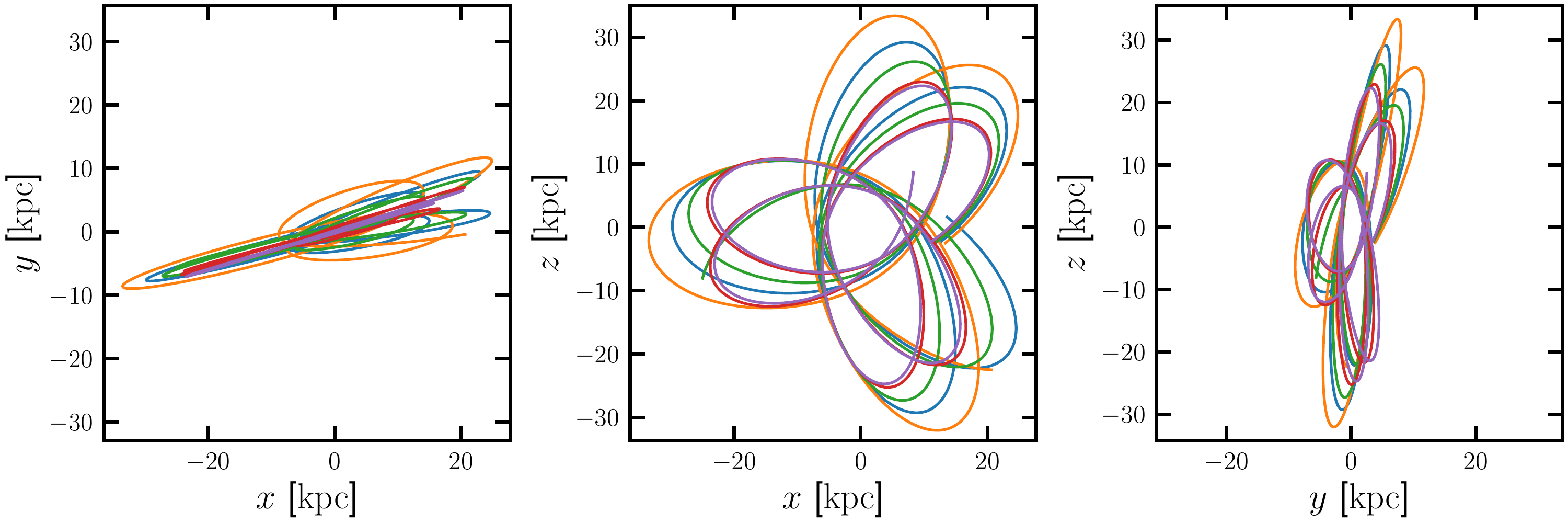}
\includegraphics[width=\textwidth]{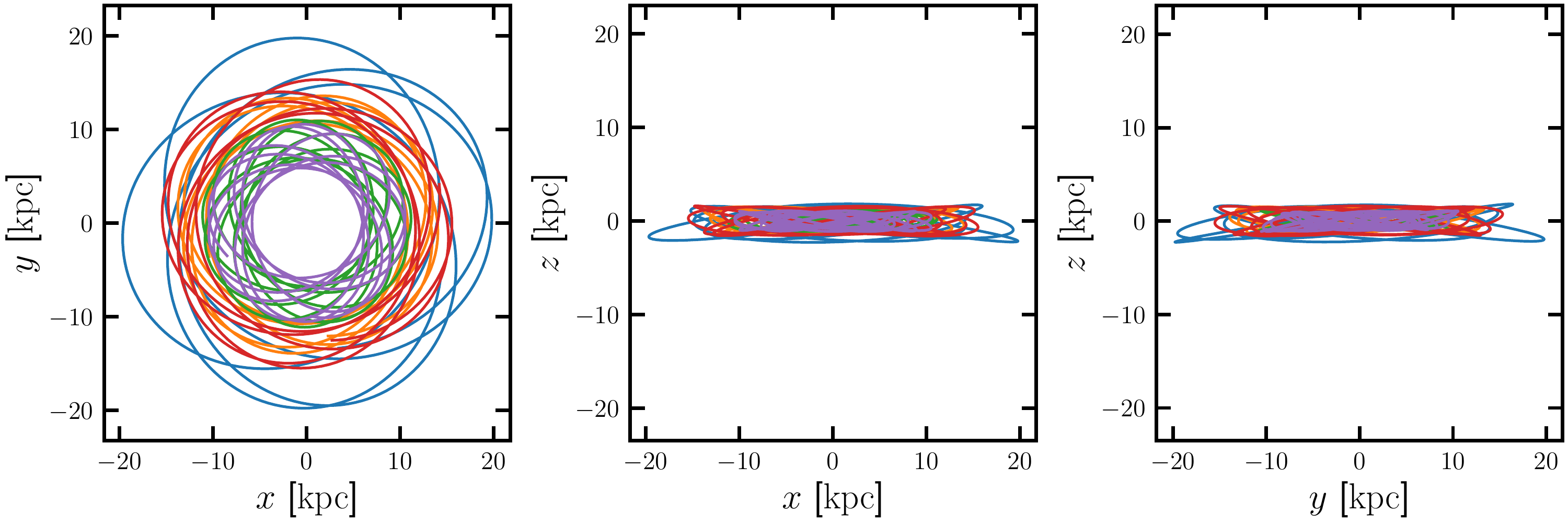}
\caption{Five example orbits of \kgotwo (top), \kgofour (middle), and \kgotwotwo (bottom) drawn from the observational uncertainties and integrated for 1~Gyr (\kgotwo) or 2~Gyr (\kgofour and \kgotwotwo). 
}
\label{fig:orbits_gc}
\end{figure*}

\begin{figure*}
\includegraphics[width=\textwidth]{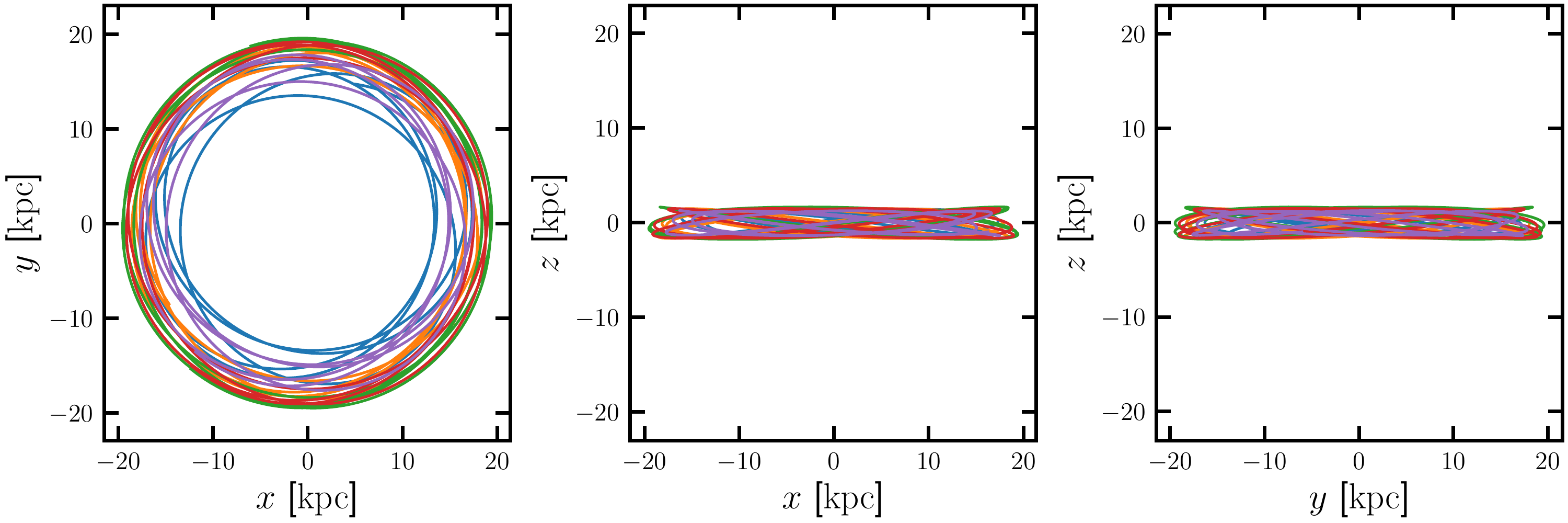}
\includegraphics[width=\textwidth]{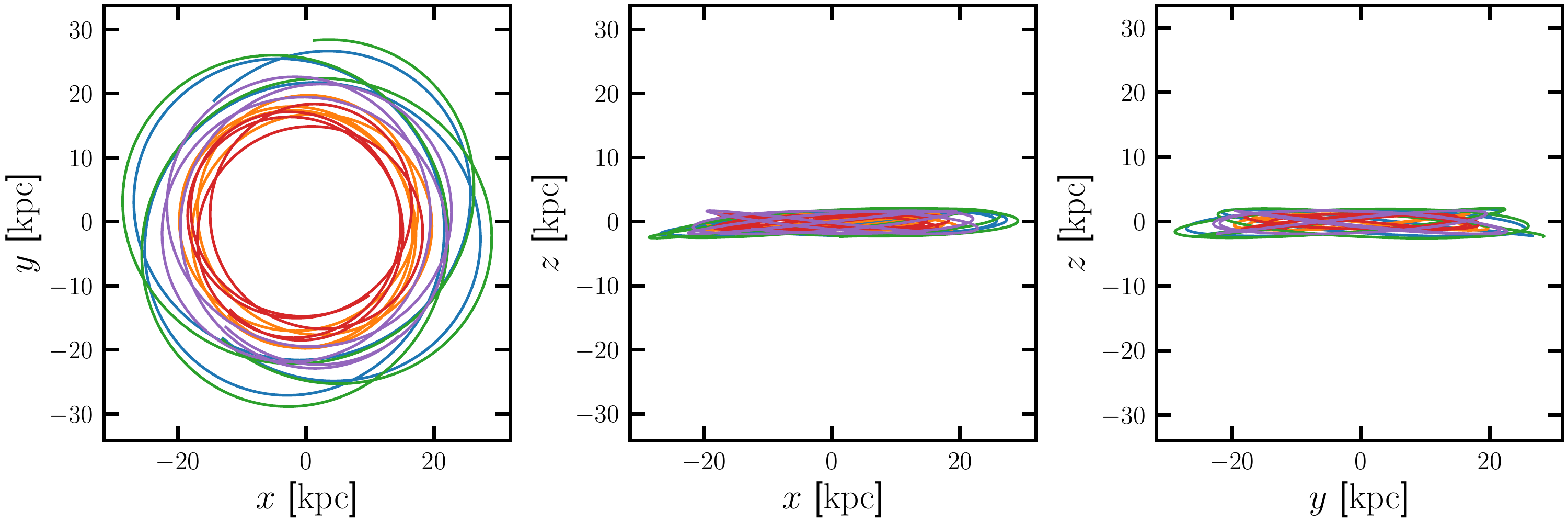}
\includegraphics[width=\textwidth]{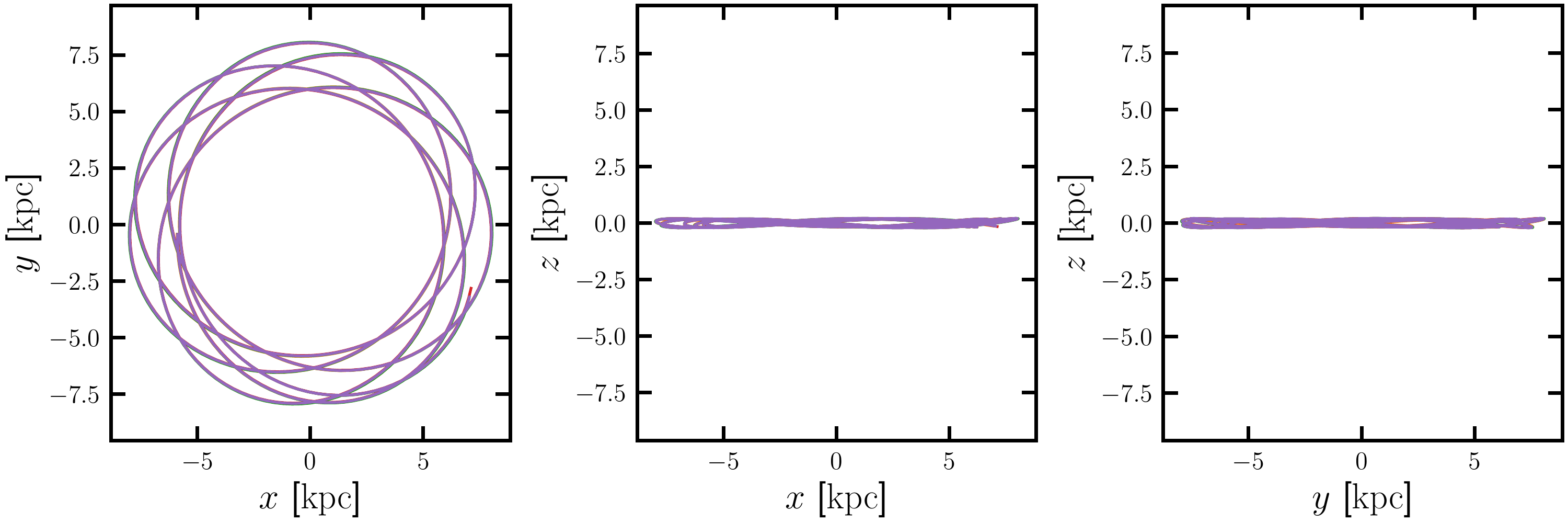}
\caption{Five example orbits of \kgoseven (top), \kgoten (middle), \kgoeight (bottom) drawn from the observational uncertainties integrated for 2~Gyr (\kgoseven and \kgoten) or 1~Gyr (\kgoeight).
}
\label{fig:orbits_opencluster}
\end{figure*}

We use the \texttt{gala} package to compute the orbits of the six star clusters and compare them to other MW globular clusters.
We use the default \texttt{MilkyWayPotential} Galactic potential from \citet{gala}.
This potential consists of two \citet{Hernquist1990ApJ...356..359H} spheroids to model the stellar bulge and nucleus, a \citet{Miyamoto1975PASJ...27..533M} axisymmetric stellar disk, and a NFW dark matter halo \citep{Navarro1996ApJ...462..563N}.  
For each cluster we compute the integrals of motion, $E$ and $L_z$, the approximately conserved quantity $L_{\perp}$ \citep{Massari2019A&A...630L...4M} and the orbital  pericenter ($r_{\rm peri}$), apocenter ($r_{\rm apo}$), and eccentricity. We list the results in Table~\ref{tab:kgo_properties}. 
We apply the same analysis to the MW globular clusters using the phase space results from \citet{Vasiliev2021MNRAS.505.5978V}. 
We compute 1000 orbits drawn from each satellite's 6D phase distribution in Table~\ref{tab:kgo_properties} and compute statistics from these runs.
We use the \texttt{astropy} v4.0 frame
\citep{Astropy2013A&A...558A..33A, Astropy2018AJ....156..123A} for the Sun's position and velocity: 
  distance to Galactic Center, $D_{\odot}= 8.122\kpc$, $v_{\odot}=(12.9, 245.6, 7.78)~\kms$
\citep{Drimmel2018RNAAS...2..210D, GravityCollaboration2018A&A...615L..15G, Reid2004ApJ...616..872R}.

In Figures~\ref{fig:orbits_gc} and ~\ref{fig:orbits_opencluster}, we show 5 example orbits drawn from the observational errors of each cluster. 
\kgotwo is on a near circular orbit ($e\sim0.12$) in the inner bulge ($R_{GC}\sim2.7\kpc$) that is confined to the plane of the disk ($z_{max}\sim 1.8\kpc$).  \kgotwo is on a retrograde orbit ($L_Z\sim +0.45~{\rm kpc^2~Myr^{-2}}$). 
\kgofour has an eccentric orbit ($e\sim0.63$) with a small pericenter ($r_{\rm peri}\sim8~\kpc$), large apocenter ($r_{\rm apo}\sim35~\kpc$), and is not confined to the MW midplane ($z_{\rm max}\sim21~\kpc$). 
\kgofour is an  halo globular cluster that is currently near passing the MW midplane.
We find that \kgotwotwo is on a circular orbit ($e\sim0.16$) that is confined to the Galactic midplane ($z_{\rm max}\sim1.3~\kpc$) at a relatively large Galactocentric radius ($\sim10-13~\kpc$). 
The orbits of the three younger open clusters are all  circular (${\rm ecc}\sim0.1-0.16$),  disk-like orbits.
Relative to  other open clusters \kgoseven and  \kgoten are at large Galactic distances ($R_{\rm GC}\sim18~{\rm and}\sim21 ~\kpc$) and have higher angular momentum $L_Z\sim -3.5 - -4.7 ~{\rm kpc^{2}~ Myr^{-2}}$.
The largest source of uncertainty in our orbit modeling comes from the distance measurement.

\section{Discussion}
\label{section:discussion}

We have presented accurate kinematics and metallicity measurements from Magellan/M2FS spectroscopy  for three recently discovered globular clusters, \kgotwo, \kgofour and \kgotwotwo, and the discovery and spectroscopic confirmation of three young open clusters, \kgoseven, \kgoten,  and \kgoeight. 
Here we consider our results in the context of the MW star cluster population.First, we comment on the nature of \kgotwotwo and whether it is an old globular cluster or open cluster (Section~\ref{sec:nature_kgo22}).  In particular, how do \kgotwo, \kgofour and \kgotwotwo relate to the MW globular cluster population and other recently discovered clusters (Section~\ref{section:gc_comparsion})? Are these new globular clusters connected to accretion events or were they formed in-situ (Section~\ref{section:gc_origin})? 
How do the open clusters compare to the Galactic radial metallicity gradient (Section~\ref{section:open_clusters})? 
We conclude by comparing our results to the literature.

\subsection{The Nature of \kgotwotwo}
\label{sec:nature_kgo22}

\citet{Garro2020A&A...642L..19G} classify \kgotwotwo as a globular cluster based on its close similarity to the globular cluster 47 Tuc but several magnitudes fainter. 
Our spectroscopic metallicity is more metal-rich ([Fe/H]$=-0.3$) than the photometric analysis ([Fe/H]$=-0.7$). The orbit of \kgotwotwo is a disk-like orbit and \kgotwotwo is confined to  the Galactic plane ($z_{\rm max}\sim 1.3~\kpc$, ${\rm ecc}\sim0.16$). Both properties are consistent with  the open cluster population.  
The age of a star cluster can be key for determining its origin as a globular cluster or open cluster \citep[e.g.,][]{Garro2022A&A...659A.155G}.

As previously noted, we had difficultly matching the spectroscopic metallicity and literature  age ($11\pm 1$ Gyr) with {\it Gaia} and DECaPS photometry as the isochrone was redder than the photometry. 
To estimate the age, we  vary the age at a fixed metallicity ([Fe/H]$=-0.3$) and check whether the color of the red-giant branch is matched. 
For this exercise, we examine both Gaia, $G_{BP}-G{RP}$, and DECaPS, $g-r$, and $r-i$ color.   
Our best estimate for the age between $2-13$ Gyr is 4 Gyr (red isochrone in Figure~\ref{fig:kgo22_summary}).  For younger ages $<3.5~{\rm Gyr}$, the main sequence turn-off would be apparent in our sample which we do not observe.  The isochrones with older ages ($>6$ Gyr) are redder than the observed data. 
In addition, we find that the {\it Gaia} RVS candidate members are better fit with the younger age.

An age of $\sim4~{\rm Gyr}$  suggests that  \kgotwotwo is an open cluster. This  agrees with the metallicity and disk-like orbit of \kgotwotwo.
An accurate age measurement from deeper photometry would confidently classify this star cluster as either an open or globular cluster.
For the reminder of the discussion, we will analyze \kgotwotwo with both the globular clusters and open clusters in our sample.

\subsection{Globular Cluster Kinematics}

\begin{figure*}
\includegraphics[width=\textwidth]{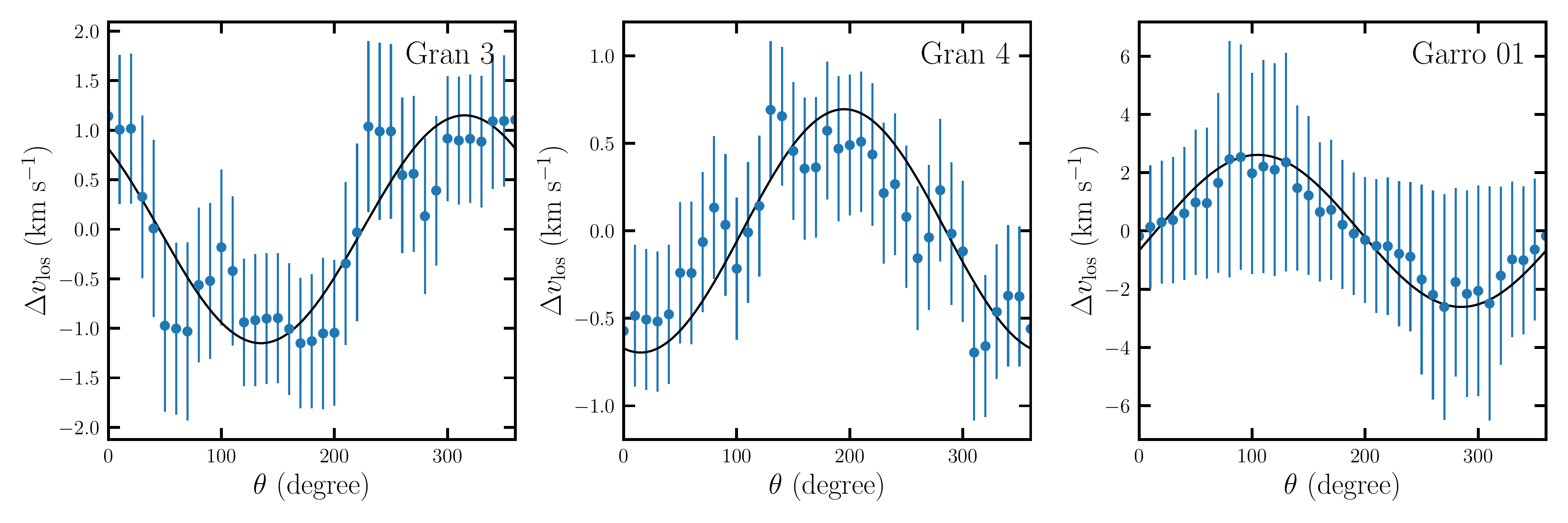}
\caption{ Rotation curves with sinusoidal by-hand fits for \kgotwo,  \kgofour, and \kgotwotwo.  
}
\label{fig:rotation_profile}
\end{figure*}

\begin{figure*}
\includegraphics[width=\textwidth]{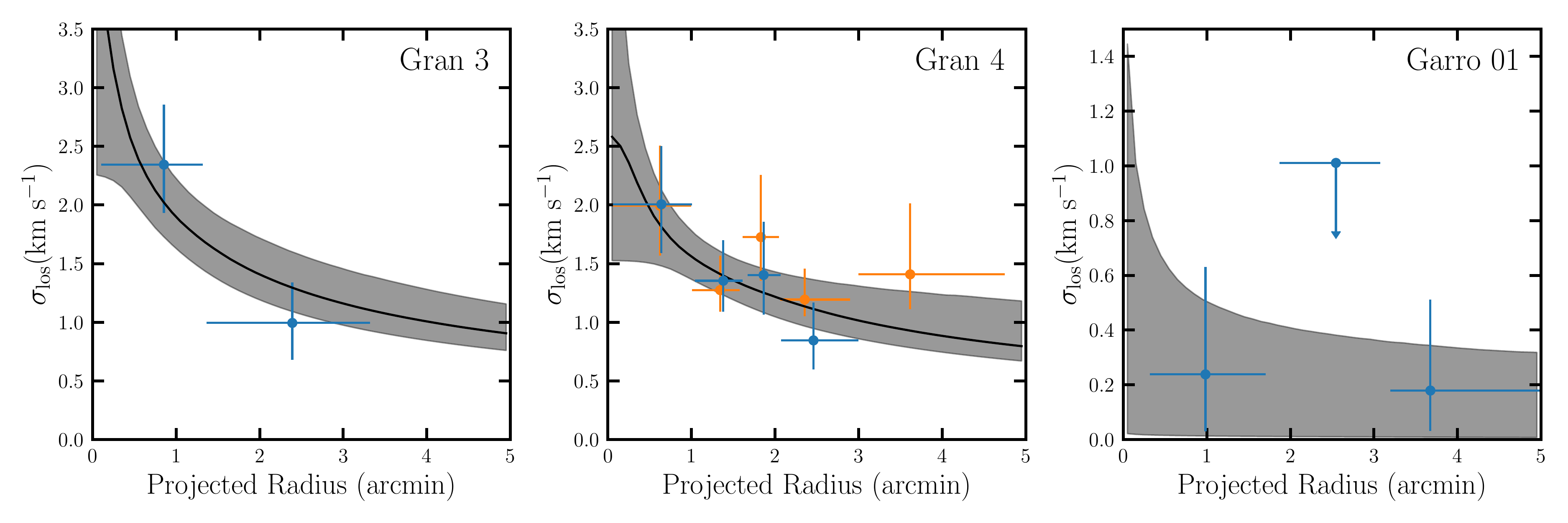}
\caption{ Projected radius versus velocity dispersion ($\sigma_{\rm los}$) profile for  \kgotwo (left), \kgofour (center), and \kgotwotwo (right). The spatial errorbar corresponds to the size of the bin and the bins have an equal number of stars (excluding the last bin). The clusters have 17, 16, and 15 stars per bin. We include velocity dispersion model fits with black lines and the shaded bands correspond to the error.
For \kgofour (center) we include a combined M2FS and AAT binned profile (orange bins). 
}
\label{fig:sigma_profile}
\end{figure*}

Globular clusters have more complex kinematics than the simple constant velocity dispersion model we have explored including rotation \citep[e.g.,][]{Sollima2019MNRAS.485.1460S} and velocity dispersion profiles \citep[][]{Baumgardt2018MNRAS.478.1520B}. 
We search for rotation by comparing the difference between the mean line-of-sight velocity across different position angles. We show the results of this exercise in Figure~\ref{fig:rotation_profile}.  There is potential  rotation on the order of $\sim1-2\kms$ in the three clusters, however, when considering the mean velocity errors it is not significant.
 
In general, globular clusters have velocity dispersion profiles that decrease with radius \citep[e.g.,][]{Baumgardt2018MNRAS.478.1520B} and we search for a radial dependence in the velocity dispersion by binning the data. 
We show the binned velocity dispersion profiles of the three globular clusters in Figure~\ref{fig:sigma_profile}. Each bin contains 18  (\kgotwo), 16 (\kgofour), and 15 stars (\kgotwotwo). For \kgofour we show the results with the M2FS data (blue) and combined M2FS + AAT data\footnote{There is a $-1.36\kms$ offset applied to the AAT data based on the repeat measurements.} (orange). Combining the AAT and M2FS data only increases the sample by 12  stars but there are 10 stars with improved velocity precision due to multiple measurements. The M2FS velocity dispersion profiles of \kgotwo and \kgofour clearly decrease with radius.  The combined M2FS + AAT profile of \kgofour is more consistent with the constant dispersion model but the central  bin has a larger velocity dispersion.
With \kgotwotwo the binned profile only measures an upper limit, similar to the global fit in our mixture model.

To better constrain and quantify the velocity dispersion profile and/or rotation we explore detailed models. 
We model the velocity dispersion with a Plummer profile \citep{Plummer1911MNRAS..71..460P} and the radially dependent velocity dispersion profile is: $\sigma^2(R) = \sigma_0^2/\sqrt{1+(r/r_0)^2}$, where $\sigma_0$, and $r_0$ are free parameters. 
We use the following rotation profile \citep[e.g.,][]{Mackey2013ApJ...762...65M, Cordero2017MNRAS.465.3515C, AlfaroCuello2020ApJ...892...20A}: $V_{\rm rot} = \frac{2 V_{\rm max} }{r_{\rm peak}}\frac{X_{PA}}{1 + \left( X_{PA}/r_{\rm peak}\right)^2}$, where $V_{\rm max}$, $r_{\rm peak}$, and $\theta_{PA}$ (which determines $X_{PA}$) are free parameters.
We use the \texttt{dynesty} package to sample the posterior and compute Bayesian evidence for model comparison \citep{Speagle2020MNRAS.493.3132S, koposov_2022_7215695_dynesty}.

We apply  the velocity dispersion and rotation models both separately and together for all three clusters. 
The inferred velocity dispersion profile fits are included in Figure~\ref{fig:sigma_profile} with their corresponding errors.
For \kgotwotwo we examine the stars with membership $>0.9$. 
For all three clusters the  functional velocity dispersion profile is favored over the constant, non-rotating model but it is not favored at a statistically significant level. 
The dispersion models for \kgotwo and \kgofour both favor a larger dispersion in the center of the clusters that decreases with radius. 
In terms of Bayesian evidence\footnote{We use the scale of \citet{Trotta2008ConPh..49...71T} to quantify significance. $\ln{Z}$ ranges of 0-1, 1-2.5, 2.5-5.5, $>5.5$ correspond to inconclusive, weak, moderate, and substantial evidence in favor of the new model.}, we find $\ln{Z}=1.7,~0.7, ~0.3$ in favor of the $\sigma(R)$ model for \kgotwo,  \kgofour, and \kgotwotwo, respectively. \kgotwo is favored with weak evidence whereas the others are inconclusive. No rotation models are favored and the rotation models produce an upper limit to the rotation. 
The rotation models do have a non-zero peak but large portions of the posterior remain consistent with no rotation. While there are coherent rotation signals in Figure~\ref{fig:rotation_profile}, the non-zero velocity dispersion and errors in the rotation curve are consistent with no rotation.
To improve constraints on the velocity dispersion profile or probe potential rotation, larger samples of stars are required.

Last, we estimate the dynamical mass and corresponding mass-to-light ratio of the three globular clusters.
Specifically, we compute the dynamical mass using the dispersion supported mass estimator from \citet{Errani2018MNRAS.481.5073E}: $M(r<1.8 R_h)\approx 3.5\times 1.8R_h \langle \sigma^2_{\rm los}\rangle G^{-1}$. 
This approximator is insensitive to the unknown underlying velocity anisotropy \citep{Walker2009ApJ...704.1274W, Wolf2010MNRAS.406.1220W, Errani2018MNRAS.481.5073E} but  assumes that the velocity dispersion is approximately constant with radius which may not be true for globular clusters or the globular clusters in our sample. 
With our line-of-sight velocity dispersion and half-light radii measurements, we measure $M(r<1.8 R_h)=2.7\times10^4 \Msun$, $4.0\times10^4 \Msun$, and $2.3\times10^3 \Msun$ ($<1.1\times10^4 \Msun$) for \kgotwo, \kgofour, and \kgotwotwo, respectively. The corresponding mass-to-light ratios are\footnote{For reference, a Plummer profile at $r=1.8 r_{\rm p}$ encloses 66.8\% of the total mass.} are 1.8 and $0.2$ ($<1.1$), for \kgofour and \kgotwotwo, respectively. There are two literature $M_V$ values for \kgotwo: $M_V=-3.8$ \citep{Garro2022A&A...659A.155G} and $M_V=-6.02$ \citep{Gran2022MNRAS.509.4962G}. These correspond to mass-to-light ratios of 1.8 (for $M_V=-6.02$) and 14.2 (for $M_V=-3.8$). 
The mass-to-light ratios for \kgotwo and \kgofour agree with old stellar populations. 
More detailed dynamical modeling, focused on star clusters would improve this analysis \citep[e.g.,][]{Gieles2015MNRAS.454..576G, Song2021MNRAS.504.4160S}.

\subsection{Comparison to the Globular Cluster Population}
\label{section:gc_comparsion}

\begin{figure*}
\includegraphics[width=\textwidth]{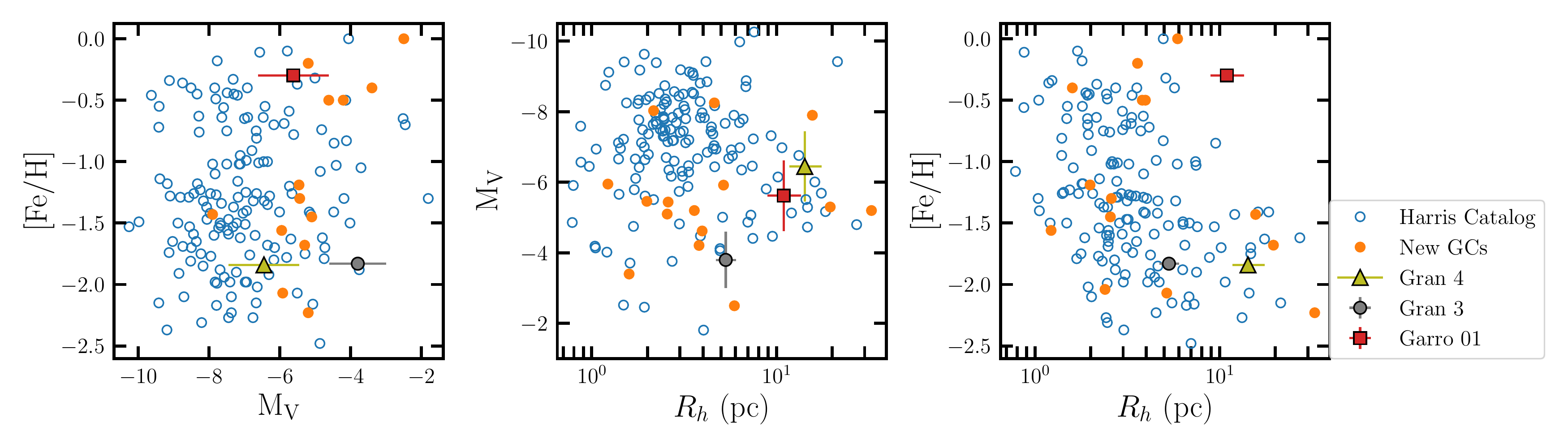}
\caption{Comparison of  \kgotwo,  \kgofour, and the ambiguous open/globular cluster \kgotwotwo with MW globular cluster population. Blue are clusters from the \citet{Harris1996AJ....112.1487H} catalog, and orange are other recently discovered globular clusters in the MW disk and bulge (see text for name and citations). 
{\bf Left:} Absolute magnitude ($M_V$) vs metallicity ([Fe/H]).
{\bf Center:} 2D half-light radius ($R_h$) vs absolute magnitude ($M_V$).
{\bf Right:} 2D half-light radius ($R_h$) vs metallicity ([Fe/H]).
}
\label{fig:gc_compare}
\end{figure*}

In Figure~\ref{fig:gc_compare}, we compare the sizes (based on the 2D projected half-light radii, $R_h$), the metallicity ([Fe/H]), and absolute magnitudes ($M_V$) of our globular cluster sample (\kgotwo, \kgofour, \kgotwotwo) to the MW globular cluster population \citep{Harris1996AJ....112.1487H} and to  other recently discovered globular clusters (or candidates) at low Galactic latitude.
The (incomplete) list of recently discovered globular clusters primarily in the Galactic disk and bulge includes: FSR~1758 \citep{Barba2019ApJ...870L..24B, Vasiliev2021MNRAS.505.5978V, Myeong2019MNRAS.488.1235M, RomeroColmenares2021A&A...652A.158R}, FSR~19 \citep{Obasi2021A&A...654A..39O}, FSR~25 \citep{Obasi2021A&A...654A..39O}, Garro~2 \citep{Garro2022A&A...662A..95G}, ESO456-29/Gran~1 \citep{Gran2019A&A...628A..45G}, Gran~2, Gran~5 \citep{Gran2022MNRAS.509.4962G}, Patchick~99 \citep{Garro2021A&A...649A..86G}, Pfleiderer~2 \citep{Ortolani2009AJ....138..889O}, Ryu~059, Ryu~879 \citep{Ryu2018ApJ...863L..38R}, VVV CL001 \citep{Minniti2011A&A...527A..81M, OlivaresCarvajal2022MNRAS.513.3993O}, VVV CL002 \citep{MoniBidin2011A&A...535A..33M}, VVV CL160 \citep{Minniti2021A&A...650L..11M}, Sagittarius~II \citep{MutluPakdil2018ApJ...863...25M, Longeard2021MNRAS.503.2754L}, and Crater~1 \citep{Weisz2016ApJ...822...32W, Kirby2015ApJ...810...56K}. We note that the classification of some objects  is uncertain and spectroscopy is needed. 

Our globular cluster sample is generally fainter than the MW globular cluster population which  explains their recent discovery  with {\it Gaia} astrometry. 
\kgofour and \kgotwotwo are both larger than most clusters ($r_h>10~\pc$).  
The large size of \kgotwotwo is particularly unusual as almost all metal-rich ([Fe/H] $>-1.5$) MW globular clusters have smaller sizes ($r_h<10~\pc$). 
The exceptions to this are Palomar 12, which is associated with Sagittarius  \citep[e.g.,][]{Law2010ApJ...718.1128L, Massari2019A&A...630L...4M}, and the Fornax~6 globular cluster associated with the Fornax dwarf spheroidal galaxy \citep{Wang2019ApJ...875L..13W, Pace2021ApJ...923...77P}. 
Unlike the other large clusters, \kgotwotwo is on a circular disk-like orbit making it less likely to have an ex-situ origin.  
\kgotwo and \kgofour are in the metal-poor tail of the MW globular cluster population as they are more metal-poor than $\sim83\%$ of the globular clusters in the \citeauthor{Harris1996AJ....112.1487H} catalog. In contrast, \kgotwotwo is one of the more metal-rich globular clusters.  If \kgotwotwo is confirmed as a younger open cluster that could explain its large size compared to other metal-rich globular clusters. 
In summary, \kgotwo, \kgofour, and \kgotwotwo have properties consistent with the  MW globular cluster population.

\subsection{Origin and Connection to Accretion Events}
\label{section:gc_origin}

\begin{figure*}
\includegraphics[width=\textwidth]{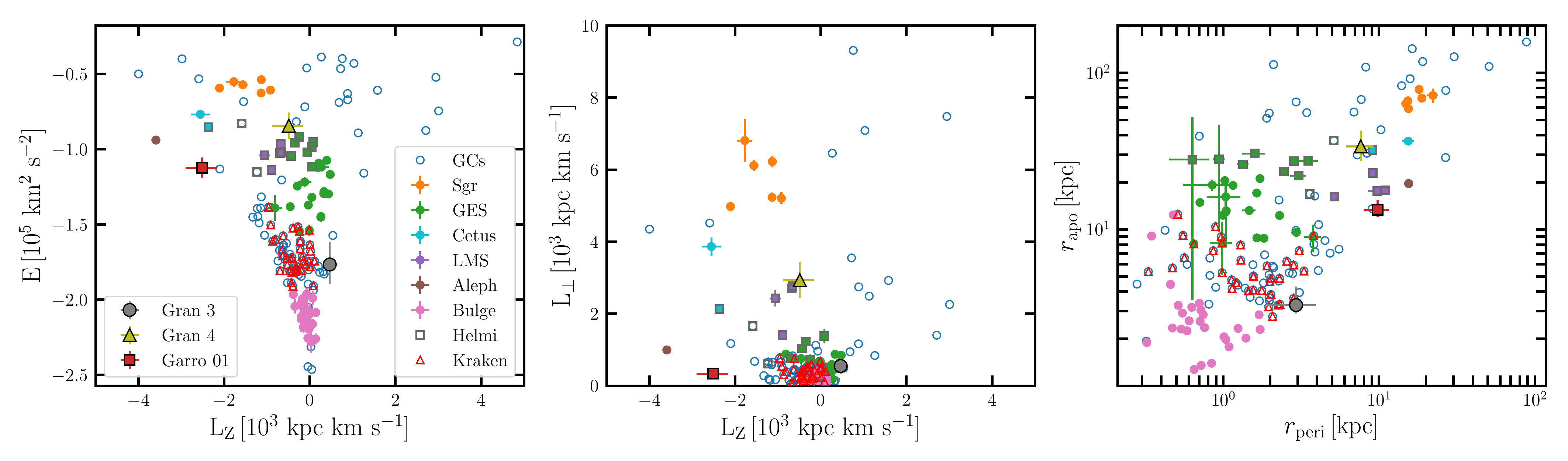}
\caption{
Comparison of \kgotwo, \kgofour, and the ambiguous open/globular cluster \kgotwotwo with the MW globular cluster population. 
{\bf Left:} energy ($E$) versus angular momentum in the z-direction ($L_z$). {\bf Center:} angular momentum in the z-direction ($L_z$) versus angular momentum in the perpendicular-direction ($L_{\perp}$). {\bf Right:} orbital pericenter ($r_{\rm peri}$) versus apocenter ($r_{\rm apo}$). The MW globular clusters are coloured according to their MW infall merger event (see text for details).
\kgofour is a candidate for the  LMS-1/Wukong merger or Helmi stream merger  and \kgotwotwo is a candidate for the  Aleph merger.
\kgotwo is a candidate member of the Galactic bulge component.
}
\label{fig:kgo_origin}
\end{figure*}

The MW has a population of in-situ and accreted/ex-situ globular clusters \citep[e.g.,][]{Myeong2018ApJ...863L..28M, Massari2019A&A...630L...4M, Kruijssen2019MNRAS.486.3180K}. 
To determine whether  \kgotwo,  \kgofour, and \kgotwotwo are associated with any accretion events we compare the orbital properties of our sample to  globular clusters associated with known events. 
In Figure~\ref{fig:kgo_origin}, we compare the orbital properties of our globular cluster sample with the MW globular cluster population. Specifically we examine the energy, angular momentum in the z-direction, and the angular momentum  in the perpendicular direction, and the pericenter and apocenter. We group the MW clusters based on accretion/merger events\footnote{We have not included the Sequoia+Arjuna+Iiloi structures \citep{Myeong2019MNRAS.488.1235M, Naidu2020ApJ...901...48N}, or Pontus structure \citep{Malhan2022ApJ...930L...9M} mergers as there is little overlap with the three globular clusters in our sample. Furthermore, Thamnos is not included as there are no known globular clusters.  }, including: Sagittarius (Sgr),  Gaia-Sausage/Enceladus (GSE) \citep{Belokurov2018MNRAS.478..611B, Helmi2018Natur.563...85H},  LMS-1/Wukong \citep{Yuan2020ApJ...898L..37Y, Naidu2020ApJ...901...48N},  Aleph \citep{Naidu2020ApJ...901...48N}, Cetus \citep{Newberg2009ApJ...700L..61N}, the Helmi stream  \citep{Helmi1999Natur.402...53H, Koppelman2019A&A...625A...5K}, low-energy/Koala/Kraken/Heracles \citep{Massari2019A&A...630L...4M, Kruijssen2019MNRAS.486.3180K, Kruijssen2020MNRAS.498.2472K, Forbes2020MNRAS.493..847F, Horta2021MNRAS.500.1385H} and the in-situ bulge population. 
We use \citet{Malhan2022ApJ...926..107M}  to associate globular clusters and merger events for GSE (we additionally include globular clusters from  \citealt{Massari2019A&A...630L...4M} for the GSE sample), Sgr, Cetus, LMS-1/Wukong, and the in-situ bulge population. 
We follow \citet{Naidu2020ApJ...901...48N} to associate  Palomar~1 and the Aleph structure.  The association of the Helmi stream globular clusters is taken from \citet{Callingham2022MNRAS.513.4107C}.  
We note between different analyses there is overlap between the globular clusters assigned to  the Helmi stream, GSE, and LMS-1/Wukong accretion events.
The  \citet{Malhan2022ApJ...930L...9M} analysis does not associate any globular clusters with the Helmi streams and the \citet{Callingham2022MNRAS.513.4107C} analysis does not  associate any globular clusters with the LMS-1/Wukong merger. 
For the Kraken merger, we use the identification from \citet{Callingham2022MNRAS.513.4107C}.  The Kraken globular clusters overlap with the low-energy globular clusters from \citet{Massari2019A&A...630L...4M}, the Heracles accretion event \citep{Horta2021MNRAS.500.1385H} and the Koala merger from \citet{Forbes2020MNRAS.493..847F} and all  may be the same merger event. For simplicity we only include the Kraken merger. 
We note the Kraken/Koala/Heracles mergers (and associated globular clusters) in terms of their chemistry are consistent with being born in-situ, in a pre-disk population known as Aurora \citep{Belokurov2022MNRAS.514..689B}.
The identification of each globular cluster with a particular merger event depends  on the methodology and sample (e.g., globular cluster, stellar stream, halo star), and different analyses have assigned the same globular cluster to different events or the in-situ population.

In the $E$-$L_Z$ plane \kgotwo is located near globular clusters associated with the Galactic bulge and the low-energy/Koala/Kraken merger \citep{Massari2019A&A...630L...4M, Kruijssen2019MNRAS.486.3180K, Kruijssen2020MNRAS.498.2472K, Forbes2020MNRAS.493..847F}. \kgotwo may be an extension the Galactic bulge component to  higher energy. 
The Kraken globular clusters generally have smaller $L_Z$ than \kgotwo and have prograde orbits with larger eccentricity \citep{Massari2019A&A...630L...4M, Kruijssen2019MNRAS.486.3180K, Kruijssen2020MNRAS.498.2472K, Forbes2020MNRAS.493..847F}. 
As this was one of the first MW mergers the globular clusters have low metallicities, similar to \kgotwo. While there is overlap, the retrograde orbit of \kgotwo disfavors an association with the Kraken merger. 
There are several retrograde accretion events in the stellar halo including the Sequoia+Arjuna+Iiloi event \citep{Malhan2022ApJ...926..107M} and Thamnos structure \citep{Koppelman2019A&A...631L...9K}. 
The Thamnos structure has the lowest energy of the retrograde structures and has a similar metallicity to \kgotwo \citep{Naidu2020ApJ...901...48N, Horta2023MNRAS.520.5671H}, however, the energy is larger than \kgotwo and it is unlikely for \kgotwo to be associated with the Thamnos merger. 
We consider \kgotwo to be a member of the Galactic bulge globular cluster group.

We find that \kgofour is closet to the LMS-1/Wukong merger event in integral of motion space ($E$, $L_Z$, $L_{\perp}$) and in orbital parameters ($r_{\rm peri}$, $r_{\rm apo}$). 
While the apocenter and energy is higher than other the clusters in LMS-1/Wukong merger, the agreement becomes better  if the distance of \kgofour decreases.
We note there is not agreement in the number or assignment of globular clusters to merger events. In particular, \citet{Callingham2022MNRAS.513.4107C} assigns the same LMS-1/Wukong globular clusters here to the Helmi streams. 
While \kgofour is close in the $E$-$L_Z$ space to globular clusters associated with the GSE merger, the GSE globular clusters have smaller $L_{\perp}$ and $r_{\rm peri}$. 
We consider \kgofour to be a candidate member of the LMS-1/Wukong merger or Helmi streams. 

\kgotwotwo has broad agreement with the energy,  angular momentum in the z-direction, eccentricity, and metallicity of the  Aleph structure \citep{Naidu2020ApJ...901...48N}. 
There is only one candidate globular cluster in this structure, Palomar~1 \citep{Naidu2020ApJ...901...48N}.
However, \kgotwotwo is confined to the disk plane ($z_{\rm max}<1.5$) and the Aleph structure has a strong vertical action and orbits with larger $z_{\rm max}$. We consider \kgotwotwo a candidate member of the Aleph structure but it is more likely to be an in-situ outer disk cluster. 

\citet{Garro2020A&A...642L..19G} suggested that \kgotwotwo could be associated with the Monoceros ring (MRi) structure \citep{Newberg2002ApJ...569..245N}. The MRi is proposed to be either the remnants of  tidally disrupted dwarf galaxy \citep[e.g.,][]{Penarrubia2005ApJ...626..128P} or a Galactic warp and flare \citep[e.g.,][]{Sheffield2018ApJ...854...47S}.  While \kgotwotwo is not near the previously detected component of  MRi, orbital analysis of MRi suggests there is overlap  at location of \kgotwotwo \citep{Conn2008MNRAS.390.1388C, Grillmair2008ApJ...689L.117G}. However, the radial velocity of the model predictions and \kgotwotwo  are offset. We measure $v_{\rm gsr}\sim-143 \kms$ for \kgotwotwo and  the radial velocity of different models varies between $v_{\rm gsr}\sim0- -100$ \citep[see Figure~17 in][]{Li2012ApJ...757..151L}. The MRi models also predict larger distances $D\gtrsim20~{\rm kpc}$ than has been inferred for the cluster and the observed metallicity distribution of MRi is more metal-poor than \kgotwotwo \citep[see Figure~13 of][]{Zhang2022ApJ...933..151Z}.  The modeling and analysis of the MRi and its connection to \kgotwotwo would benefit from a dedicated search in this region of sky but the radial velocity disagreement suggests they are not associated.

\subsection{Tracing the Galactic Metallicity Gradient with Open Clusters}
\label{section:open_clusters}

\begin{figure}
\includegraphics[width=\columnwidth]{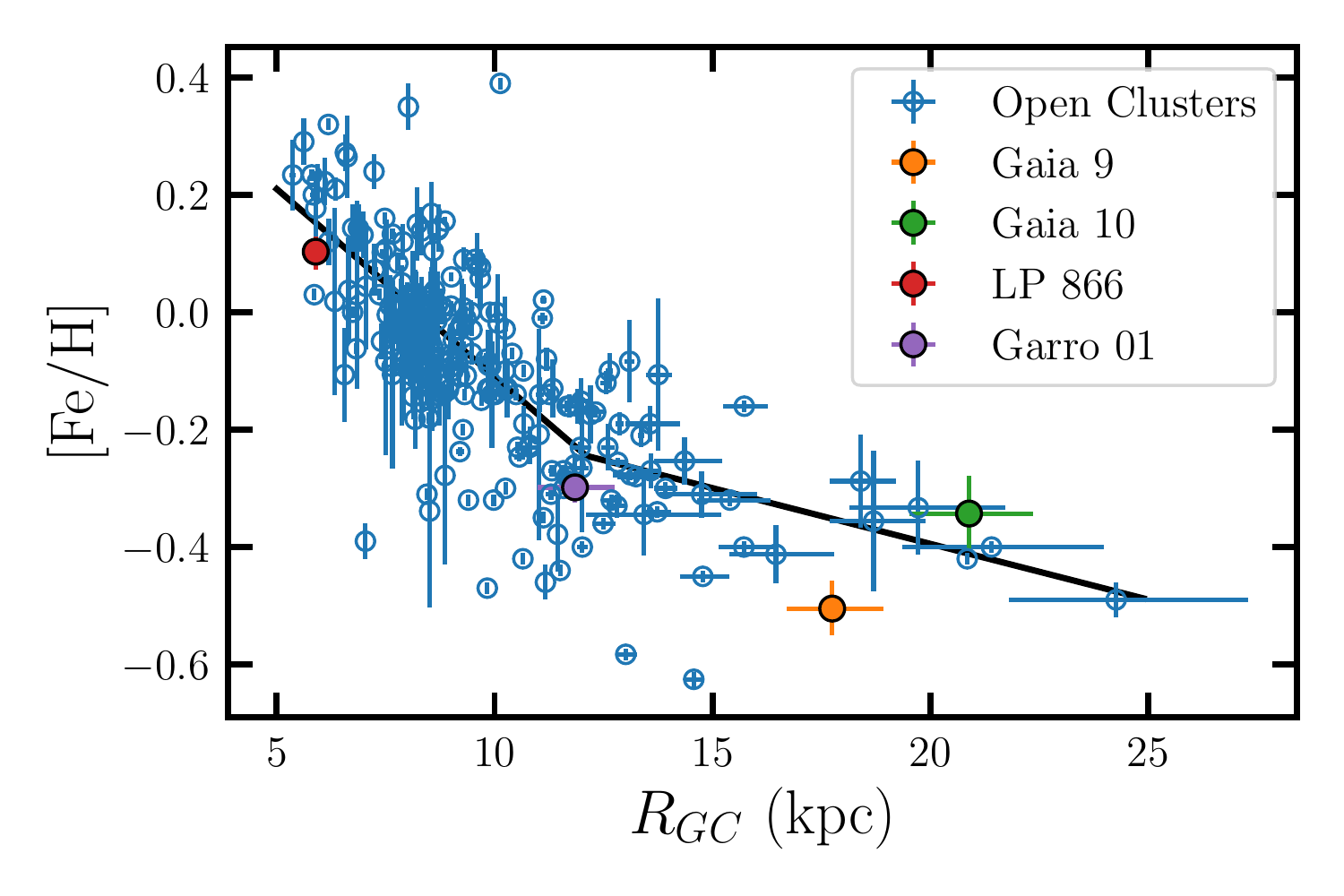}
\includegraphics[width=\columnwidth]{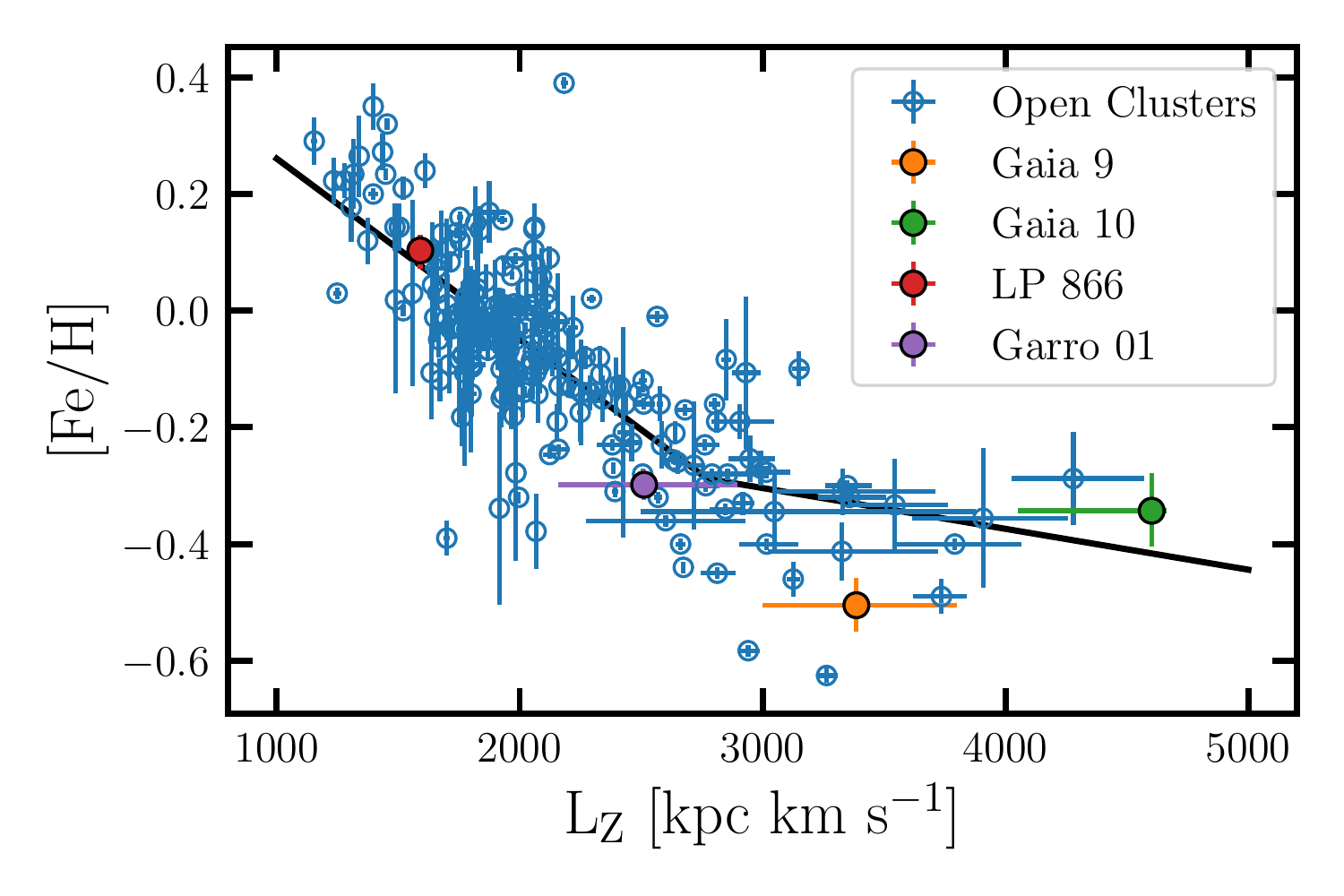}
\caption{Comparison of \kgoseven, \kgoten, \kgoeight, and the ambiguous open/globular cluster \kgotwotwo with MW open cluster population. We include literature measurements from \citet{Spina2022Univ....8...87S} as blue points.
{\bf Top}: Galactocentric radius ($R_{GC}$) versus metallicity ([Fe/H]). 
{\bf Bottom}: angular momentum in the Z-direction ($L_{Z}$) versus metallicity ([Fe/H]). 
In both panels, we include the best fit relations derived from the literature open cluster sample in  \citet{Spina2022Univ....8...87S}. 
}
\label{fig:oc_r_gc_feh}
\end{figure}

Open clusters are an important tracer of the Galactic radial metallicity gradient as each cluster can be age-dated with individual chemical elements studied \citep[e.g., ][]{Jacobson2016A&A...591A..37J, Donor2020AJ....159..199D, Spina2021MNRAS.503.3279S, Gaia_RecioBlanco2023A&A...674A..38G}. 
The Galactic metallicity gradient  traces  Galactic formation and evolution scenarios and the complex interplay between star formation, stellar evolution, stellar migration, gas flows, and cluster disruption in the Galactic disk \citep[e.g.,][]{Chiappini1997ApJ...477..765C, Schonrich2009MNRAS.396..203S, Kubryk2015A&A...580A.126K, Spitoni2019A&A...623A..60S}.

We  compare the three new open clusters (\kgoseven, \kgoten, \kgoeight)  to the literature open clusters \citep[from][]{Spina2022Univ....8...87S}  and the  Galactic radial metallicity distribution  in Figure~\ref{fig:oc_r_gc_feh}. 
The \citet{Spina2022Univ....8...87S} sample includes open cluster data from several different literature surveys including: APOGEE, Gaia-ESO, GALAH, OCCASO, and SPA. 
We include the best fit relation relation between $R_{GC}-{\rm [Fe/H]}$ and $L_{Z}-{\rm [Fe/H]}$ for literature open clusters from \citet{Spina2022Univ....8...87S}. 
We include comparisons to the Galactic radius and the the angular momentum in the z-direction ($L_Z$). $L_Z$ is conserved and the current Galactic radius may not be not representative of their birth radius as  open clusters may have undergone radial migration \citep[e.g.,][]{Chen2020MNRAS.495.2673C}.
In both cases, the best fit relation becomes shallower at large radius.  Previous measurements have suggested that the relation flattens out at large radii \citep[e.g.][]{Frinchaboy2013ApJ...777L...1F, Donor2020AJ....159..199D}.

The   open clusters analyzed here are in general agreement with the open cluster population trends with metallicity,  Galactocentric radius, and $L_Z$. 
While our spectroscopic follow-up has only measured [Fe/H] for three more open clusters, \kgoseven and \kgoten are among the most metal-poor open clusters in the MW open cluster population.   \kgoten has \NEW{the largest $L_Z$ of any MW open} cluster. 
The properties of \kgotwotwo are consistent with the Galactic metallicity gradient as traced by open clusters.
 Future analyses of the Galactic radial metallicity gradient will be improved by including \kgotwotwo, \kgoseven, and \kgoten and the  metallicity measured  from our Magellan/M2FS spectroscopy.

\subsection{Comparison to Previous Studies}
\label{sec:compare}

Of the six star clusters studied only  \kgotwo has previous spectroscopic follow-up. 
\citet{Gran2022MNRAS.509.4962G} presented VLT/MUSE spectroscopy of \kgotwo and found $v_{\rm los}=74.32 \pm 2.70\kms$ and ${\rm [Fe/H] }= -2.37 \pm 0.18$. Both measurements are discrepant with our results and other \kgotwo spectroscopic studies \citep{FernandezTrincado2022A&A...657A..84F, Garro2023A&A...669A.136G}. 
There is a $\sim20~\kms$ offset between the mean radial velocities measured in \citet{Gran2022MNRAS.509.4962G} compared to our results and literature. 
While velocity zero-point offsets of a few $\kms$ are common between different instruments/methods, this value is too large to be caused by a zero point offset. 
It is unclear what the origin of this offset is. We note that all our members are consistent with the same radial velocity, stellar parameters from a single stellar population, a single metallicity, and consistent proper motions. 
In Figure~10 of \citet{Gran2022MNRAS.509.4962G}, the proper motions are consistent with the majority of their members having similar proper motions but there are only a few radial velocity members. Some members may be missing due to the field-of-view of \kgotwo but it is possible that the radial velocity peak was misidentified.

\citet{FernandezTrincado2022A&A...657A..84F} analyzed high resolution APGOEE spectroscopy of  two stars in \kgotwo. 
Due to their sample of two stars, the mean velocity of \kgotwo they measure is offset from our by $\sim 4\kms$. 
The metallicity from \citet{FernandezTrincado2022A&A...657A..84F} is [Fe/H]$=-1.7\pm0.09$  is larger than our measurement but it is consistent within uncertainties.
\citet{Garro2023A&A...669A.136G} identified 6 members in the {\it Gaia} RVS sample and their velocity measurement ($v_{\rm los}=93.1 \pm 3.6 \kms$) agrees with our measurement within  uncertainties. 
In our analysis of the {\it Gaia} RVS data we identified one additional \kgotwo member (Section~\ref{sec:kgo2}).

We find larger angular sizes for \kgotwo and \kgofour compared to \citet{Gran2022MNRAS.509.4962G}.
For reference, \citet{Gran2022MNRAS.509.4962G} find $R_h = 1.05 \pm 0.04~{\rm arcmin}$ and $R_h = 1.14 \pm 0.02~{\rm arcmin}$  for \kgotwo and \kgofour, respectively, compared to our values of $R_h = 1.7 \pm 0.2~{\rm arcmin}$ and $R_h = 2.2_{-0.4}^{+0.5}~{\rm arcmin}$ for \kgotwo and \kgofour, respectively.  The source of this discrepancy could be due to different photometry ({\it Gaia} versus near-IR) or fitting methodology. 
For  \kgotwotwo there is excellent agreement between our King profile fits ($r_c=1.8_{-0.5}^{+0.7}\arcmin$) and the results ($r_c=2.1\pm1.5\arcmin$) in \citet{Garro2020A&A...642L..19G}. 
We note that the absolute magnitude of \kgotwo is discrepant between  \citet{Garro2022A&A...659A.155G} ($M_V \sim -3.8$) and  \citet{Gran2022MNRAS.509.4962G} ($M_V\sim -6.02$).

Our orbital analysis of \kgotwo is similar to literature results \citep{Gran2022MNRAS.509.4962G, FernandezTrincado2022A&A...657A..84F, Garro2023A&A...669A.136G}. 
Both \citet{FernandezTrincado2022A&A...657A..84F} and \citet{Garro2023A&A...669A.136G} include a rotating bar in their modeling which is not included in our modeling. 
Compared to the other studies the value of $z_{\rm max}$ is smaller and the energy lower. We attribute this to the lower distance we assumed in this work. Compared to \citet{FernandezTrincado2022A&A...657A..84F} and \citet{Garro2023A&A...669A.136G} we have a more circular orbit (lower eccentricity) which agrees with \citet{Gran2022MNRAS.509.4962G}.
We note that compared to other studies we have  more precise  velocity and proper motion measurements.

\section{Conclusion }
\label{section:conclusion}

We have presented the spectroscopic follow-up of three recently discovered globular clusters and three recently discovered open clusters.
Our main findings are as follows:

\begin{itemize}

\item We have independently discovered three globular clusters (\kgotwo /Patchick~125, \kgofour, \kgotwotwo) and three open clusters (\kgoseven, \kgoten, \kgoeight) with {\it Gaia} astrometry. \kgoseven and \kgoten are new discoveries presented here.

\item We have presented spectroscopic follow-up with Magellan/M2FS and measured stellar parameters of  601 stars and identified 273 members across 6 star clusters and confirmed the legitimacy of all six clusters. 
In addition, we have presented AAT/AAOmega spectroscopy of \kgofour which confirms our M2FS results.

\item We find  \kgotwo (Patchick~125) is an old, metal-poor  globular cluster on a retrograde orbit trapped within the  Galactic bulge. 
From our M2FS spectroscopy, we identified 37 members and measured a heliocentric velocity of $v_{\rm los}=90.9 \pm0.4 \kms$ and  metallicity of ${\rm [Fe/H]}=-1.83_{-0.03}^{+0.04}$. 
In addition, there are 2 APOGEE and 7 {\it Gaia} RVS members. 
From our orbital analysis, \kgotwo has a near circular orbit (${\rm ecc}\sim0.07$) and orbital pericenter and apocenter of $2.9~{\rm kpc}$ and $3.3~{\rm kpc}$, respectively.
\kgotwo is likely an in-situ bulge globular cluster.

\item  \kgofour is an old, metal-poor globular cluster with a halo-like orbit that is passing though the Galactic mid-plane. 
We identified 63 members from our M2FS spectroscopy and 22 members (12 unique) from our AAT/AAOmega spectroscopy. 
We measured a heliocentric velocity of $v_{\rm los}=-266.4\pm0.2\kms$ and  metallicity of ${\rm [Fe/H]}=-1.84\pm0.02$. 
In addition, there are 3 {\it Gaia} RVS members. 
From our orbital analysis, \kgofour has an eccentric orbit circular orbit (${\rm ecc}\sim0.63$) and orbital pericenter and apocenter of $7.6~{\rm kpc}$ and $33.9~{\rm kpc}$, respectively.
\kgofour is a candidate member of the LMS-1/Wukong and/or Helmi stream merger events. 

\item  \kgotwotwo is a metal-rich star cluster on an outer disk-like orbit. We identified 43 members with our M2FS spectroscopy and measured a heliocentric velocity of $v_{\rm los}=31.0\pm0.1\kms$ and  metallicity of ${\rm [Fe/H]}=-0.30\pm0.03$.  There is more overlap in velocity with the MW foreground and we constructed a mixture model to quantitatively account for  the MW foreground.  
In addition, there are 2 candidate {\it Gaia} RVS members.
We found that \kgotwotwo has a relatively large size ($R_h\sim 11~{\rm pc}$) compared to other metal-rich globular clusters ($R_h<5~{\rm pc}$). 
From our orbital analysis, \kgotwotwo has a  circular orbit (${\rm ecc}\sim0.16$) and orbital pericenter and apocenter of $9.8~{\rm kpc}$ and $13.3~{\rm kpc}$, respectively. 
We estimated an age of 4 Gyr, which is younger than previous analysis \citep[11$\pm0.5$ Gyr][]{Garro2020A&A...642L..19G}. Combined with the metallicity and orbit, this suggests that \kgotwotwo is an open cluster but a confident classification requires a more detailed age measurement and we consider the classification ambiguous. 

\item Both \kgotwo and \kgofour have evidence for radially declining velocity dispersion profiles (Figure~\ref{fig:sigma_profile}). There is inconclusive evidence for rotation in the globular clusters (Figure~\ref{fig:rotation_profile}). 

\item We have confirmed \kgoseven, \kgoten, and \kgoeight as open clusters and identified 19-83 spectroscopic members from our M2FS spectroscopy. 
We measured metallicities of $-0.50$, $-0.34$, and $+0.10$ and estimated ages of 1.5, 1, and 3 Gyr from isochrone fits for \kgoseven, \kgoten, and \kgoeight, respectively.
All three open clusters are on circular, disk-like orbits. 
\kgoseven and  \kgoten are among the most distant ($R_{GC}\sim18,~21.2~\kpc$) and  most metal-poor open clusters known and have some of the largest angular momentum in the z-direction.
These clusters will assist in tracing the Galactic metallicity gradient to larger radii (Figure~\ref{fig:oc_r_gc_feh}).

\end{itemize}

The Milky Way star cluster population remains incomplete and  {\it Gaia}  astrometry has revolutionised our understanding of star clusters. We have spectroscopically confirmed six star clusters and there remain many more candidate star clusters that require spectroscopic follow-up.

\section*{Acknowledgements}

We thank the referee for their helpful comments.
ABP is supported by NSF grant AST-1813881.  M.G.W. acknowledges support from NSF grants AST-1813881 and AST-1909584. SK  was partially supported by NSF grants AST-1813881 and AST-1909584.
EO was partially supported by NSF grant AST-1815767.
NC is supported by NSF grant AST-1812461.
MM was supported by U.S.\ National Science Foundation (NSF) grants AST-1312997, AST-1726457 and AST-1815403.
IUR acknowledges support from NSF grants AST-1613536, AST-1815403, AST-2205847, and PHY-1430152 (Physics Frontier Center/JINA-CEE).
DBZ acknowledges support from Australian Research Council grant DP220102254.
We thank Lorenzo Spina for sharing their open cluster catalog. 
EO wants to remember Jill Bechtold here.

For the purpose of open access, the author has applied a Creative Commons
Attribution (CC BY) licence to any Author Accepted Manuscript version arising
from this submission.

This work has made use of data from the European Space Agency (ESA)
mission {\it Gaia} (\url{https://www.cosmos.esa.int/gaia}), processed by the {\it Gaia} Data Processing and Analysis Consortium (DPAC,
\url{https://www.cosmos.esa.int/web/gaia/dpac/consortium}). Funding
for the DPAC has been provided by national institutions, in particular
the institutions participating in the {\it Gaia} Multilateral Agreement.

This work has used data acquired at the Anglo-Australian Telescope. We acknowledge the traditional custodians of the land on which the AAT stands, the Gamilaraay people, and pay our respects to elders past and present.

This research has made use of NASA's Astrophysics Data System Bibliographic Services.
This paper made use of the Whole Sky Database (wsdb) created by Sergey Koposov and maintained at the Institute of Astronomy, Cambridge by Sergey Koposov, Vasily Belokurov and Wyn Evans with financial support from the Science \& Technology Facilities Council (STFC) and the European Research Council (ERC).

Funding for the Sloan Digital Sky 
Survey IV has been provided by the 
Alfred P. Sloan Foundation, the U.S. 
Department of Energy Office of 
Science, and the Participating 
Institutions. 

SDSS-IV acknowledges support and 
resources from the Center for High 
Performance Computing  at the 
University of Utah. The SDSS 
website is www.sdss4.org.

SDSS-IV is managed by the 
Astrophysical Research Consortium 
for the Participating Institutions 
of the SDSS Collaboration including 
the Brazilian Participation Group, 
the Carnegie Institution for Science, 
Carnegie Mellon University, Center for 
Astrophysics | Harvard \& 
Smithsonian, the Chilean Participation 
Group, the French Participation Group, 
Instituto de Astrof\'isica de 
Canarias, The Johns Hopkins 
University, Kavli Institute for the 
Physics and Mathematics of the 
Universe (IPMU) / University of 
Tokyo, the Korean Participation Group, 
Lawrence Berkeley National Laboratory, 
Leibniz Institut f\"ur Astrophysik 
Potsdam (AIP),  Max-Planck-Institut 
f\"ur Astronomie (MPIA Heidelberg), 
Max-Planck-Institut f\"ur 
Astrophysik (MPA Garching), 
Max-Planck-Institut f\"ur 
Extraterrestrische Physik (MPE), 
National Astronomical Observatories of 
China, New Mexico State University, 
New York University, University of 
Notre Dame, Observat\'ario 
Nacional / MCTI, The Ohio State 
University, Pennsylvania State 
University, Shanghai 
Astronomical Observatory, United 
Kingdom Participation Group, 
Universidad Nacional Aut\'onoma 
de M\'exico, University of Arizona, 
University of Colorado Boulder, 
University of Oxford, University of 
Portsmouth, University of Utah, 
University of Virginia, University 
of Washington, University of 
Wisconsin, Vanderbilt University, 
and Yale University.

Software: \texttt{dynesty} \citep{Speagle2020MNRAS.493.3132S, koposov_2022_7215695_dynesty}, \texttt{astropy} \citep{Astropy2013A&A...558A..33A, Astropy2018AJ....156..123A},
\texttt{matplotlib} \citep{matplotlib}, 
\texttt{NumPy} \citep{numpy},
\texttt{iPython} \citep{ipython},
\texttt{SciPy} \citep{2020SciPy-NMeth},
\texttt{corner.py} \citep{corner}, 
\texttt{emcee} \citep{ForemanMackey2013PASP..125..306F} ,
\texttt{Q3C} \citep{2006ASPC..351..735K}, 
\texttt{gala} \citep{gala}, \texttt{galpy} \citep{Bovy2015ApJS..216...29B}, \texttt{MultiNest} \citep{Feroz2008MNRAS.384..449F, Feroz2009MNRAS.398.1601F}.

\section*{Data Availability}

We provide our Magellan/M2FS and AAT/AAOmega catalogs and a machine readable version of Table~\ref{tab:kgo_properties} at Zenodo under a Creative Commons Attribution license: \href{https://doi.org/10.5281/zenodo.7809128}{doi:10.5281/zenodo.7809128}.
The other catalogs used in our analysis ({\it Gaia} DR3, DECaPS, APOGEE) are publicly available.


\bibliographystyle{mnras}
\bibliography{main_bib_file, extra_bib} 





\bsp	
\label{lastpage}
\end{document}